\documentclass[11pt]{article}

\usepackage{amsmath}
\usepackage{amssymb}
\usepackage{fullpage}
\usepackage{graphicx}
\usepackage{bm}% bold math
\usepackage{hyperref}% add hypertext capabilities
\usepackage{cite}
\usepackage{authblk}
\usepackage{lineno}
\usepackage{xcolor}
\PassOptionsToPackage{hyphens}{url}\usepackage{hyperref}

\title{\textbf{The evolution of punishing institutions}}
\date{}

\author{\textbf{Mohammad Salahshour}\thanks{\texttt{salahshour.mohammad@gmail.com}.}}
\affil{Max Planck Institute for Mathematics in the Sciences, Inselstrasse 22, D-04103, Leipzig, Germany}

\begin{document}
\maketitle

%\linenumbers

\section*{Abstract}
A large body of empirical evidence suggests that humans are willing to engage in costly punishment of defectors in public goods games. Based on such pieces of evidence, it is suggested that punishment serves an important role in promoting cooperation in humans, and possibly other species. Nevertheless, theoretical work has been unable to show how this is possible. The problem originates from the fact that punishment, being costly, is an altruistic act and its evolution is subject to the same problem that it tries to address. To suppress this so-called second-order free-rider problem, known theoretical models on the evolution of punishment resort to one of the few established mechanisms for the evolution of cooperation. This leaves the question that whether altruistic punishment can evolve and give rise to the evolution of cooperation, unaddressed. Here, by considering a population of individuals who play a public goods game, followed by a public punishing game, introduced here, we show that altruistic punishment indeed evolves and promotes cooperation, in a general environment and in the absence of a cooperation favoring mechanism. Besides, our analysis shows, being close to a physical phase transition facilitates the evolution of altruistic punishment.

%\linenumbers

\section*{Introduction}
Cooperation requires a cooperator to incur a cost for others to benefit. As such, cooperation is costly and expected to diminish by natural selection \cite{Axelrod,Nowak}. Contrary to this rational expectation, cooperation is everywhere-present in the biological and social world \cite{West,Clutton,RandDavid,Fehr0}. Empirical studies suggest, by enforcing cooperation in animal \cite{Tibbetts,Hauser0,Wenseleers,Ratnieks} and human societies \cite{Fehr1,Fehr2,Gachter,Gurerk,Fehr0,Mathew}, altruistic punishment can play an important role in the evolution of cooperation. However, just as cooperation does, altruistic punishment, being costly, goes against an individual's self-interest, and its evolution is yet another puzzle \cite{Milinski,Panchanathan,Dreber}. Many studies have tried to address this so-called second-order free-rider problem, according to which, free-riding on social punishers who punish defectors, results in the elimination of punishers, and subsequently, to the extinction of cooperators by first-order free-riding on cooperators. Theoretical models have been able to show that these two problems, first-order and second-order free-rider problems, can simultaneously be solved, in cooperation favoring environments; that is when one of the few known mechanisms for the evolution of cooperation is at work. In this regard, group selection \cite{Boyd}, indirect reciprocity \cite{Sigmund,Panchanathan,Hilbe}, voluntary participation \cite{Fowler,Hauert,Sigmund2,Garcia}, and the spatial selection and network structure \cite{Nakamaru,Brandt,Szolnoki}, have been successfully appealed to show how cooperation and punishment can co-evolve in cooperation favoring environments. However, the important question that whether altruistic punishment on its own can promote cooperation has remained unaddressed.

The theoretical grounds appear even more disappointing, when it is noticed that in many cooperation favoring environments, where it is argued that punishment and cooperation can co-evolve, the inclusion of a complete set of possible strategies, by adding antisocial punishers (who defect and punish cooperators) to the population, leads to the invasion of antisocial punishers, and thus, undermines the co-evolution of social punishment and cooperation \cite{Rand1,Rand2,Hauser}. A priory, there is no reason why antisocial punishment should be excluded in the model. Instead, especially given the empirical evidence that anti-social punishment is abundant in human and animal societies \cite{Herrmann,Denant,Gachter2,Nikiforakis}, its exclusion is a point which a proper theory needs to address. Although this problem is solved in some cases of a cooperation favoring environments (for example, in the case of structured populations \cite{Szolnoki}, when a reputation mechanism is at work \cite{Hilbe}, or when participation is voluntary and the type of institutions are observable \cite{Garcia}), a satisfactory understanding of the extent to which anti-social punishment can prevent the evolution of social behavior is still lacking. These two problems, the second-order free-rider problem and, to a lesser degree, the antisocial punishment problem, raise important questions about the evolution of altruistic punishment and its role in the evolution of cooperation, not only in a general environment but also in many cooperation favoring environments.

Here, by considering a well-mixed population of individuals who play a public goods game (PGG), followed by a public punishing game, introduced here, we show that the first-order and the second-order free-rider problems can be solved simultaneously in a general environment, that is, in absence of any cooperation favoring mechanism and in the presence of antisocial punishment. This establishes altruistic punishment as a fundamental road to the evolution of cooperation and explains its evolution. Furthermore, by considering the same model in a structured population, we show that the mechanism is further strengthened in a cooperation favoring environment. Besides, we provide evidence that being close to a physical phase transition facilitates the evolution of social punishment. We argue how the public punishing game admits an intuitive interpretation in terms of the law enforcing institutions commonly observed in human societies \cite{Ostrom,Veszteg}. In this regard, we argue that increasing the adaptivity of the model, in a way that it more closely resembles human punishing institutions, can make the co-evolution of cooperation and social punishment possible even in more hostile conditions for the evolution of cooperation.

\section*{The Model}
To see how altruistic punishment can evolve and promote cooperation, we consider a population of $N$ individuals in which groups of $g$ individuals are formed at random to play a public goods game. This game is frequently appealed in studies on the evolution of cooperation \cite{Fehr1,Fehr2,Gachter,Gurerk,Milinski,Dreber,Salahshour0,Salahshour00}. In this game, each individual can cooperate or defect. Cooperators pay a cost $c$ to invest in a public resource. Defectors invest nothing. All the investments are multiplied by an enhancement factor $r$ and are divided equally among all the group members. In addition to playing the PGG, individuals can engage in social or antisocial punishment. For this purpose, cooperators can invest an amount $c'$ in a social punishment pool. In the same way, defectors can invest the same amount $c'$ in an antisocial punishment pool. All the investments in the social and antisocial punishment pools are multiplied by a punishment enhancement factor $\rho$ and are used for punishment purposes. To this goal, a fraction $1-\alpha$ of the total resources in the social punishment pool is spent to punish defectors in the group, and a fraction $\alpha$ is used to punish cooperators who do not contribute to the social punishment pool. In the same way, a fraction $1-\alpha$ of the total resources in the antisocial punishment pool is spent to punish cooperators, while a fraction $\alpha$ is used to punish defectors who do not contribute to the antisocial punishment pool.

Individuals gather payoff according to the payoff structure of the game and reproduce with a probability proportional to the exponential of their payoff such that the population size remains constant. That is, each individual in the next generation is offspring to an individual in the last generation with a probability proportional to the exponential of its payoff. Offspring inherit the strategies of their parent subject to mutations. We assume mutations in the decisions of the individuals to contribute to the public pool, and their decision to contribute to the punishment pool occurs independently, each with probability $\nu$. In this study, we set $c=c'=1$.

\begin{figure}
	\includegraphics[width=\linewidth, trim = 50 255 50 20, clip,]{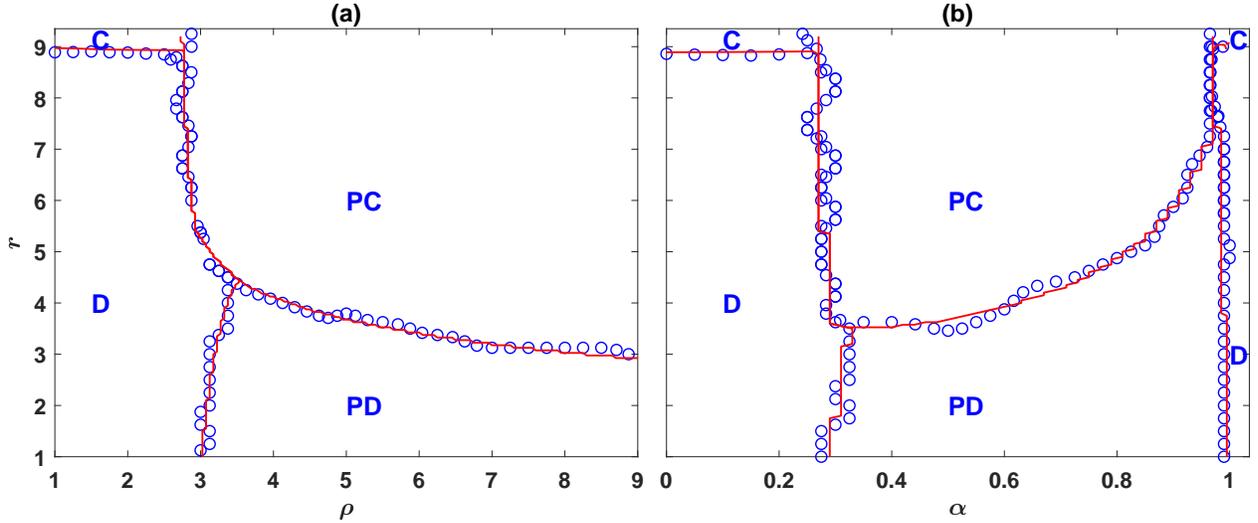}
	\caption{The phase diagram of the model in the case of a well-mixed population. Blue circles denote the results of a simulation in a population of size $N=10000$, and the red lines denote the results of the replicator dynamics. Depending on the parameters of the model, the model shows four different phases. $C$, $D$, $PC$, and $PD$ denote different phases in which, respectively, cooperators, defectors, punishing cooperators, and punishing defectors dominate. Here, $g=9$, $\nu=0.001$, $c=c'=1$. In (a) $\alpha=0.5$ and in (b) $\rho=5$.}
	\label{figure1g9}
\end{figure}
\section*{Results}
The phase diagrams of the model in the $r-\rho$ and $r-\alpha$ planes are presented in, respectively, Fig. (\ref{figure1g9}.a) and Fig. (\ref{figure1g9}.b). The blue circles present the results of simulations in a population of size $N=10000$, and the red lines result from the numerical solutions of the replicator dynamics, developed in the Methods section. To drive the phase diagram, we have determined the equilibrium state of the system starting from a random initial condition in which the strategies of the individuals are randomly assigned. For the replicator dynamics, this amounts to an initial condition in which the frequency of all the strategies is the same.

Depending on the parameters of the model, the system can be found in one of the four possible phases. Beginning with the phase diagram in the $r-\rho$ plane, as can be seen in Fig. (\ref{figure1g9}.a), for small punishment enhancement factors $\rho$, punishing strategies do not evolve. In this region, for $r$ smaller than the group size $g=9$, the population settles into a defective phase in which only non-punishing defectors survive. This phase is denoted by $D$ in the figures. As $r$ increases, for a value of $r$ close to $g=9$, a phase transition to a phase where non-punishing cooperators survive occurs. This phase is denoted by $C$. On the other hand, for large values of $\rho$, punishing strategies evolve and eliminate other strategies. However, the nature of the evolving punishment, depends on the value of $r$. For small $r$ and large values of $\rho$, such that the return to the investment in the public pool is low, but that to the investment in the punishment pool is high, the population settles into the antisocial punishment phase, where antisocial punishers dominate the population. This phase is indicated by $PD$. On the other hand, for large enough values of $r$ and $\rho$, such that the returns to investments in both the public pool and the punishment pool are high, the dynamics settle into the social punishment phase, where social punishers dominate the population. This phase is denoted by $PC$. 

\begin{figure}%[!ht]
	\includegraphics[width=\linewidth, trim = 50 255 50 20, clip,]{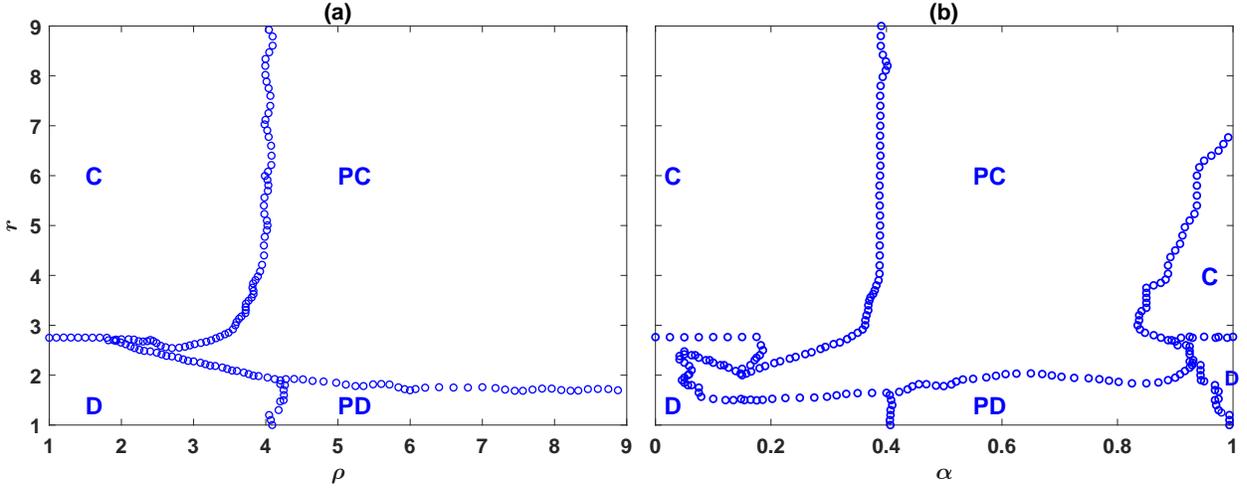}
	\caption{The phase diagram of the model in a structured population. The phase diagram is derived by running simulations in a population of size $40000$, residing on a $200\times 200$ first nearest neighbor lattice with Moore connectivity and periodic boundaries. Depending on the parameters of the model, the model shows four different phases. $C$, $D$, $PC$, and $PD$ denote different phases in which, respectively, cooperators, defectors, punishing cooperators, and punishing defectors dominate. Here, $g=9$, $\nu=0.001$, and $c=c'=1$. In (a) $\alpha=0.5$ and in (b) $\rho=5$.}
	\label{fignet1}
\end{figure}
The phase diagram in the $r-\alpha$ plane shows similar phases. For small $\alpha$, such that there is not enough investment in punishing second-order free riders, punishment does not evolve. In this region, for $r$ smaller than a value close to $g=9$ non-punishing defectors survive, and for larger values of $r$ non-punishing cooperators survive. As $\alpha$ increases, a discontinuous transition occurs above which punishing strategies evolve. In this regime, for small $r$ punishing defectors dominate. However, for larger values of $r$, punishing cooperators dominate. For very large values of $\alpha$ (close to $1$), another transition occurs above which punishment does not evolve. This shows enough investment in punishing first-order free riders is also necessary for punishing institutions to evolve.
Interestingly, for larger values of $\alpha$, the evolution of social punishment requires a larger value of $r$. This shows, an optimal value of $\alpha$, that is, an optimal weight of second-order with respect to first-order punishment exists which facilitates the evolution of social punishment. Altogether, our analysis reveals that for punishing institutions to evolve, they need to punish both first-order free riders, who do not contribute to the public good, and second-order free riders, who do not contribute to the punishment pool. This suggests, for instance, issuing a fine for non-contributors to the policing institutions is necessary for the evolution of such institutions.

\begin{figure*}[!ht]
	\includegraphics[width=\linewidth, trim = 75 260 30 50, clip,]{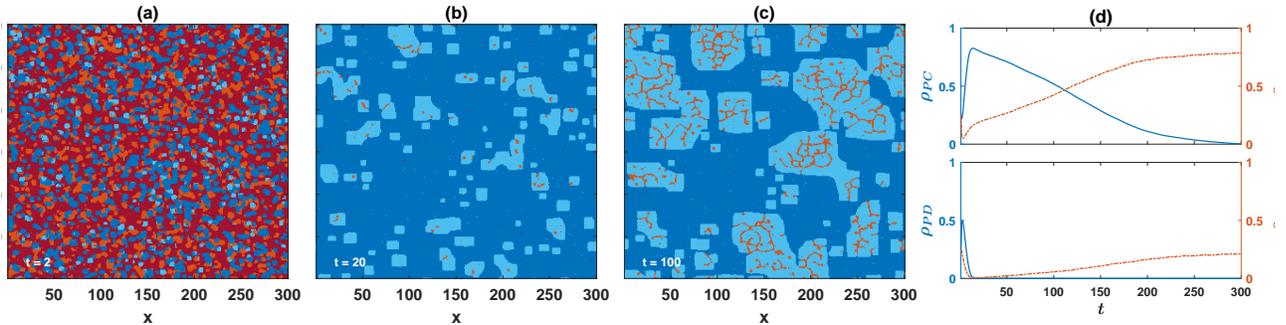}
	\caption{Time evolution of the model in a structured population. (a) to (c) present the snapshots of the time evolution of the system, and (d) shows the density of different strategies as a function of time. Here, a population of $N=90000$ individuals live on a two dimensional $300\times300$ lattice with periodic boundaries and Moore connectivity. Different strategies are indicated by different colors. Light blue shows non-punishing cooperators, dark blue shows punishing cooperators, light red shows non-punishing defectors, and dark red shows punishing defectors. The simulation is started with random initial condition. Here, $\nu=10^{-3}$, $c=c'=1$, $r=4$ and $\rho=3.47$.}
	\label{fig3}
\end{figure*}
So far we have considered a well-mixed population. As population structure favors the evolution of cooperation, one might expect that social punishment to evolve in the case of structured populations as well. To see this is indeed the case, we present the phase diagram of the model for a structured population, in Fig. (\ref{fignet1}.a) and Fig. (\ref{fignet1}.b). The phase diagram is derived by performing simulations in a population of $N=40000$ individuals residing on a $200\times200$ first nearest neighbor square lattice with Moore connectivity and periodic boundaries. In the case of a structured population, the model shows similar phases to those that appeared for a well-mixed population. However, two shifts are observable in the position of the phase transitions. First, due to network reciprocity, the $D-C$ transition shifts to smaller enhancement factors. We note that, in the absence of punishing strategies, the $D-C$ transition for the same network structure and size occurs for a larger value of $r$. This shows the beneficial effect of the introduction of the punishing strategies for the evolution of cooperation, even in the parameter regimes where such punishing strategies do not evolve. This interesting phenomenon results from a rock-paper-scissor like dynamics, according to which, punishing cooperators facilitate the evolution of non-punishing cooperators by eliminating defectors and facilitating the formation of small cooperators block (See S.4.4 and S.5). Second, away from the $D-C$ transition, the phase transition from non-punishing to punishing strategies shifts to larger values of $\rho$, compared to the mixed population. This shows, surprisingly, population structure can hinder the evolution of punishing institutions. In addition, the analysis of the model reveals, the evolution of social punishment is facilitated close to the $D-C$ transition. This can be observed to be the case by noting that close to the $D-C$ transition, the value of $\rho$ above which social punishment evolves decreases. As shown in the Supplemental Material (SM) (S.4.2), this result holds for other parameter values and shows the beneficial effect of being close to a continuous transition for the evolution of social behavior. Finally, we note that the $PD-PC$ transition occurs for a smaller value of $r$ compared to the $D-C$ transition. Furthermore, by increasing $\rho$, the $PD-PC$ transition shifts to smaller values of $r$. This is the case in both a well-mixed population and a structured population, and shows, the more effective the punishment, the easier and for smaller enhancement factors, social punishment and cooperation evolve.

To take a closer look at the mechanism by which cooperation and punishment co-evolve, in Fig. (\ref{fig3}.a) to  Fig. (\ref{fig3}.c), we present the snapshots of the time evolution of the system close to the $C-PC$ phase transition. Here, a population of $N=90000$ individuals, residing on a $300\times300$ lattice with Moore connectivity and periodic boundaries, is considered. The simulation starts with a random assignment of the strategies. The density of different strategies as a function of time is plotted in Fig. (\ref{fig3}.d). Starting from a random initial condition, punishing strategies rapidly grow, while the non-punishing strategies decline. Punishing defectors have the highest growth rate at the beginning of the simulation. This results in a sharp increase in their number by driving other solitary strategies into extinction. However, after small blocks of punishing cooperators are formed, they obtain the highest growth by reaping the benefit of cooperation among themselves and avoiding being punished by rival punishing defectors, and rapidly drive punishing and non-punishing defectors into extinction. As argued below, by setting the stage for the invasion of non-punishing cooperators, this phenomenon facilitates the evolution of cooperation. The initial rapid growth of punishing cooperators sets the stage for the second stage of the time evolution of the system, in which small domains of non-punishing cooperators are formed in a sea of social punishers. As here, the system is in the $C$ phase, cooperators experience an advantage with respect to social punishers. Consequently, cooperators blocks start to grow slowly along the horizontal and vertical boundaries, until they dominate the population. While defectors can not survive in the sea of the social punishers, they do survive by forming narrow bands within the domain of non-punishing cooperators. Consequently, once non-punishing cooperators start to dominate the population, the density of defectors increases as well. We note that, as shown in the SM (S.4.4 and S.5), and the Supplementary Videos, this coarsening pattern is characteristic of the evolution of punishing strategies.

Finally, we note, in a mixed population the model is multi-stable in the entire phase diagram; for $r<g$, all the three strategies, $D$, $PD$, and $PC$ are stable. This implies that all the transitions but the $D-C$ transition are discontinuous. As shown in the SM (S.3.2), the nature of the $D-C$ transition depends on the value of $\rho$. While this transition is discontinuous for large $\rho$, for small $\rho$ there is a cross over from the $D$ phase to the $C$ phase without passing any singularity. In between, the transition becomes a continuous transition at a critical point. Similarly, for a structured population, all the transitions but the $D-C$ transition are discontinuous. The $D-C$ transition, in contrast, shows no discontinuity and appears to occur continuously (S.4.3).

\section*{Discussion}
As the evolution of altruistic punishment is riddled by the same kind of free-riding problem that the evolution of cooperation is, it was believed that resorting to another cooperation favoring mechanism is necessary to explain the evolution of altruistic punishment and its role in the evolution of cooperation \cite{Boyd,Sigmund,Panchanathan,Fowler,Hauert,Sigmund2,Nakamaru,Brandt,Szolnoki}. As we have shown, this is not necessarily the case. Instead, the efficient coupling of the second-order and first-order free-rider problems provides a surprising way for the simultaneous solution of both dilemmas. This establishes altruistic punishment as a fundamental road to the evolution of cooperation and can explain its overwhelming presence in human and many animal societies. Furthermore, our study brings new insights into the beneficial conditions for the evolution of punishing institutions and social behavior. In this regard, our analysis shows there exists an optimal weight for second-order, with respect to first-order punishment which facilitates the evolution of altruistic punishment. Besides, the more efficient the punishment mechanism, the more likely that social punishment as opposed to anti-social punishment evolves. It also reveals network structure can be detrimental to the evolution of punishing institutions. This theoretical prediction has been observed recently in spatial public goods experiments \cite{Li}, and parallels some arguments that network structure can sometimes be surprisingly harmful for the evolution of social behavior \cite{Hauert2}. Finally, we have seen that being close to a physical phase transition is beneficial for the evolution of punishing institutions. This parallels many arguments that being close to physical phase transitions can provide optimal conditions for many biological functions, and extends such arguments to the evolution of social behavior \cite{Salahshour,Bialek,Mora,Hidalgo}.

Just as the public goods game is thought of as a metaphor for a social dilemma, the public punishing game introduced here, can be thought of as a metaphor for public punishing institutions, such as formal and informal policing institutes at work in human societies \cite{Ostrom,Veszteg}. In this regard, the contribution to the public punishing pool can be thought of as a tax paid by individuals to establish a policing institute. Similarly, the punishment of second-order free-riders can be considered as a fine for not paying the tax, and the punishment of first-order free-riders can be considered as fine for not contributing to the public good.
Our model can be thought of as a simple and minimal model which grasps the essential aspects of the evolution of such punishing institutions. However, human punishing institutions, are adaptive institutions which have accumulated a high level of sophistication in the course of their evolution \cite{Ostrom,Veszteg}. In terms of this analogy, the model can be made more adaptive to more closely resemble human sanctioning institutions. Such adaptivity is expected to increase the effectiveness of the punishing institutions, and thus facilitates the evolution of cooperation and social punishment, just as it arguably does in real-world sanctioning institutions. In the SM, we consider one such modification. In the model considered so far, if there is nobody to punish, the resources in the punishing pool are wasted. This, while keeping the model simple, might not be realistic. In real-world punishing institutions, wealth is not destroyed. Instead, if not necessary for sanctioning purposes, it can be used for other purposes, such as charity or reward. In the SM (S.1.3, S.2, and S.6), we consider such a non-wasteful punishment model and show that, as intuitively expected, such a modification indeed facilitates the evolution of cooperation and social punishment.

\section*{Materials and Methods}

\subsection*{The replicator dynamics}
In a well-mixed population, it is possible to drive a set of equations for the time evolution of the system in terms of the replicator-mutation equations. In a general case, the replicator-mutation equations can be written as follows:
\begin{align}
m_{x}(t+1)=\sum_{x'}\nu_{x}^{x'}m_{x'}(t)\frac{w_{x'}(t)}{\sum_{x''}m_{x''}(t)w_{x''}(t)}.
\label{eqrepMM}
\end{align}
Here, $x$, $x'$, and $x''$ refer to the strategies, and can be $PC$, $C$, $PD$, or $D$, referring to, respectively, punishing cooperators, non-punishing cooperators, punishing defectors, and non-punishing defectors. $m_x$ is the density of the strategy $x$, $w_{x}$ is the expected fitness of an individual with strategy $x$, and $\nu_{x}^{x'}$ is the mutation rate from the strategy $x'$ to the strategy $x$. Under our assumption that mutations in the strategies of the individuals in the public goods pool, and in the public punishing pool occur independently, these can be written in terms of the probability of mutation $\nu$, as follows. For those transformations which require no mutations, that is $x=x'$, we have $\nu_{x}^{x'}=1-2\nu+\nu^2$ (this is the probability that no mutation, neither in the strategy to contribute to the public pool, nor in the strategy to contribute to the punishing pool, occurs). For those  rates which require two mutations, one in the decision of the individuals to contribute to the public pool, and one in their decision to contribute to the punishing pool, we have $\nu_D^{PC}=\nu_C^{PD}=\nu_{PC}^D=\nu_{PD}^C=\nu^2$. All the other rates, which require only one mutation, are equal to $\nu_C^D=\nu_D^C=\nu_{PC}^{PD}=\nu_{PD}^{PC}=\nu_C^{PC}=\nu_{PC}^C=\nu_{PD}^D=\nu_{D}^{PD}=\nu-\nu^2$.

To use the replicator-mutation equation, eq. \eqref{eqrepMM}, we need expressions for the expected fitness of different strategies. These are given by the following equations:
\begin{align}
w_{PC}=&\sum_{n_{PD}=0}^{g-1-n_{PC}-n_{C}}\sum_{n_{C}=0}^{g-1-n_{PC}}\sum_{n_{PC}=0}^{g-1}\exp\bigg[r\frac{1+n_{C}+n_{PC}}{g} -(1-\alpha)\rho\frac{n_{PD}}{1+n_{PC}+n_{C}}-c-c'\bigg]\nonumber\\&{\rho_{PC}}^{n_{PC}}{\rho_{C}}^{n_{C}}{\rho_{PD}}^{n_{PD}}{\rho_{D}}^{g-1-n_{PC}-n_{C}-n_{PD}}\binom{g-1}{n_{PC},n_{C},n_{PD},g-1-n_{PC}-n_{C}-n_{PD}},\nonumber
\end{align}
\begin{align}
w_{C}=&\sum_{n_{PD}=0}^{g-1-n_{PC}-n_{C}}\sum_{n_{C}=0}^{g-1-n_{PC}}\sum_{n_{PC}=0}^{g-1}\exp\bigg[r\frac{1+n_{C}+n_{PC}}{g} -(1-\alpha)\rho\frac{n_{PD}}{1+n_{PC}+n_{C}}-\alpha\rho\frac{n_{PC}}{1+n_{C}}-c\bigg]\nonumber\\&{\rho_{PC}}^{n_{PC}}{\rho_{C}}^{n_{C}}{\rho_{PD}}^{n_{PD}}{\rho_{D}}^{g-1-n_{PC}-n_{C}-n_{PD}}\binom{g-1}{n_{PC},n_{C},n_{PD},g-1-n_{PC}-n_{C}-n_{PD}},\nonumber\\
w_{PD}=&\sum_{n_{PD}=0}^{g-1-n_{PC}-n_{C}}\sum_{n_{C}=0}^{g-1-n_{PC}}\sum_{n_{PC}=0}^{g-1}\exp\bigg[r\frac{n_{C}+n_{PC}}{g} -(1-\alpha)\rho\frac{n_{PC}}{1+n_{PD}+n_{D}}-c'\bigg]\nonumber\\&{\rho_{PC}}^{n_{PC}}{\rho_{C}}^{n_{C}}{\rho_{PD}}^{n_{PD}}{\rho_{D}}^{g-1-n_{PC}-n_{C}-n_{PD}}\binom{g-1}{n_{PC},n_{C},n_{PD},g-1-n_{PC}-n_{C}-n_{PD}},\nonumber\\
w_{D}=&\sum_{n_{PD}=0}^{g-1-n_{PC}-n_{C}}\sum_{n_{C}=0}^{g-1-n_{PC}}\sum_{n_{PC}=0}^{g-1}\exp\bigg[r\frac{n_{C}+n_{PC}}{g} -(1-\alpha)\rho\frac{n_{PC}}{1+n_{PD}+n_{D}}-\alpha\rho\frac{n_{PD}}{1+n_{D}}\bigg]\nonumber\\&{\rho_{PC}}^{n_{PC}}{\rho_{C}}^{n_{C}}{\rho_{PD}}^{n_{PD}}{\rho_{D}}^{g-1-n_{PC}-n_{C}-n_{PD}}\binom{g-1}{n_{PC},n_{C},n_{PD},g-1-n_{PC}-n_{C}-n_{PD}},\nonumber\\
\label{eqpayoffM}
\end{align}
In the following we explain how these expressions can be derived. In the process, we consider a focal individual in a group where there are $n_{PC}$ punishing cooperators, $n_C$ non-punishing cooperators, $n_{PD}$ punishing defectors, and $n_D=g-1-n_{PC}-n_C-n_{PD}$ non-punishing defectors in the group. The term in the large bracket in eq. (\ref{eqpayoffM}), is the payoff of such a focal individual. In the following, we explain why this is so.

Using the previously mentioned notation for the group composition of a focal individual, $r\frac{1+n_{PC}+n_C}{g}-c$ is the payoff of a focal punishing or non-punishing cooperator, and $r\frac{n_{PC}+n_C}{g}$ is the payoff of a focal punishing or non-punishing defector from the public goods game. These are the first terms in the large bracket in eq. \eqref{eqpayoffM}. A focal punishing or non-punishing cooperator, receives a punishment from the punishing defectors in its group equal to $(1-\alpha)\rho\frac{n_{PD}}{1+n_{PC}+n_{C}}$. This is the second term in the large bracket in the expressions for $w_{PC}$ and $w_C$. In addition, a focal non-punishing cooperator is punished by punishing cooperators in its group, by an amount equal to $\alpha\rho\frac{n_{PC}}{1+n_{C}}$. This is the third term in the large bracket in the expression for $w_C$. In the same way, a focal punishing or non punishing defector, receives a punishment from the punishing cooperators in its group equal to $(1-\alpha)\rho\frac{n_{PC}}{1+n_{PD}+n_{D}}$. This is the second term in the large bracket in the expressions for $w_{PD}$ and $w_D$. In addition, a focal non-punishing defector is punished by punishing defectors in its group, by an amount equal to $\alpha\rho\frac{n_{PD}}{1+n_{D}}$. This is the third term in the large bracket in the expression for $w_D$. Finally, as punishing cooperators contribute to both the public pool and the social punishing pool, they pay a cost of $c+c'$. On the other hand, non-punishing cooperators only pay a cost of $c$ to contribute to the public pool. Similarly, punishing defectors, pay a cost of $c'$ to contribute to the antisocial punishing pool. Defectors contribute to none of the pools and pay no cost.

As individuals reproduce with a probability proportional to their payoff, the expected fitness of a strategy can be defined as the expected value of the exponential of the payoff of that strategy. To calculate the expected value of fitness, we note that ${\rho_{PC}}^{n_{PC}}{\rho_{C}}^{n_{C}}{\rho_{PD}}^{n_{PD}}{\rho_{D}}^{g-1-n_{PC}-n_{C}-n_{PD}}$ $\binom{g-1}{n_{PC},n_{C},n_{PD},g-1-n_{PC}-n_{C}-n_{PD}}$, is the probability that a focal individual finds itself in a group with $n_{PC}$ punishing cooperators, $n_{C}$ non-punishing cooperators, $n_{PD}$ punishing defectors, and $n_{D}$ non-punishing defectors. Here, $\binom{g-1}{n_{PC},n_{C},n_{PD},g-1-n_{PC}-n_{C}-n_{PD}}=\frac{(g-1)!}{n_{PC}!,n_{C}!,n_{PD}!,(g-1-n_{PC}-n_{C}-n_{PD})!}$ is the multinational coefficient. This is the number of ways that among the $g-1$ group-mates of a focal individual, $n_{PC}$, $n_{C}$, $n_{PD}$, and $g-1-n_{PC}-n_{C}-n_{PD}$ individuals are respectively, punishing cooperators, non-punishing cooperators, punishing defectors, and non-punishing defectors. Summation over all the possible configurations gives the expected fitness of different strategies. Using the expressions in eq. (\ref{eqpayoffM}) for the expected fitness of different strategies in eq. (\ref{eqrepMM}), we have a set of four equations which gives an analytical description of the model, in the limit of infinite population size. 

\subsection*{The simulations and analytical solutions}
Analytical solutions result from numerically solving the replicator dynamics of the model. Simulations of the model are performed according to the model definition. Both simulations and analytical solutions are performed with an initial condition in which all the strategies are found in similar frequencies in the population pool. For the solutions of the replicator dynamics, this is assured by setting the initial frequency of all the four strategies equal to $1/4$. For simulations, this is assured by a random assignment of the strategies.

\section*{Acknowledgment} The author acknowledges funding from Alexander von Humboldt Foundation in the framework of the Sofja Kovalevskaja Award endowed by the German Federal Ministry of Education and Research.

\nolinenumbers

\setcounter{figure}{0}
\setcounter{equation}{0}
\setcounter{section}{0}

\newcommand{\red}{\textcolor{red}}
\newcommand{\blue}{\textcolor{blue}}
\newcommand{\Boydthree}{24}
\newcommand{\Hanaki}{25}
\newcommand{\figmixed}{1}
\newcommand{\fignet}{2}
\newcommand{\eqpayoff}{2}

\renewcommand\thesection{S. \arabic{section}}
\renewcommand\thesubsection{\thesection.\arabic{subsection}}

\renewcommand\thefigure{S.\arabic{figure}} 
\renewcommand\thetable{S.\arabic{table}} 
\renewcommand\theequation{S.\arabic{equation}} 
\makeatletter
\def\p@subsection{}
\makeatother

\clearpage
\onecolumn

\begin{center}
	{\huge \bf 
		
		Supplementary Material for:}
	\\
	{\LARGE\bf	The Evolution of Punishing Institutions\\ }
	\vspace{0.4cm}
	
	Mohammad Salahshour
	%	\\
	%XXXXX	
	%	{\small \it Max Planck Institute for Mathematics in the Sciences, Leipzig, Germany}
	%{\small \it XXXX}
	
\end{center}

\section{Overview of the Model}
We consider two population structures: a well-mixed population, and a structured population. In the following, we bring the model description for the two different population structures, separately. In addition, we will present an adaptive or non-wasteful punishment model in which if there is nobody to punish, the resources in the punishing pool is redistributed among contributors to the punishing pool.
\subsection{Well-mixed population}
In the case of a well-mixed population structure, we consider a population of $N$ individuals. At each time step, groups of $g$ individuals are formed at random to play a public goods game (PGG), followed by a public punishing game. In the PGG stage, each individual can either cooperate to defect. Cooperators pay a cost $c$ to invest the same amount $c$ in a public pool. Defectors pay no cost and invest nothing in the public pool. All the investments in the public pool are multiplied by an enhancement factor $r$ and are divided equally among the individuals in the group. After playing the PGG, individuals play a public punishing game. In this stage, cooperators can pay a cost $c'$ to contribute to a social punishing pool, or they can refrain from contributing to the social punishing pool. In the same way, defectors can pay a cost $c'$ to contribute to an antisocial punishing pool, or they can refrain from contributing to the anti-social punishing pool. We call cooperators who contribute to the social punishing pool, punishing cooperators, and those who do not, are called non-punishing cooperators (occasionally both are called cooperators for short). Similarly, defectors who contribute to the antisocial punishing pool are called punishing defectors, and those who do not, are called non-punishing defectors (occasionally both are called defectors for short). All the contributions to a punishing pool are multiplied by an enhancement factor $\rho$ and are spent for punishment purposes. To this end, we assume a fraction $\alpha$ of the resources in the social punishing pool is spent to punish cooperators who do not contribute to the social punishing pool (such that each non-punishing cooperator in a group with $n_C$ non-punishing cooperators and $n_{PC}$ punishing cooperators is punished by an amount $\alpha\rho n_{PC}/n_C)$), and the rest, a fraction $1-\alpha$, is spent to punish (punishing and non-punishing) defectors (such that in a group with $n_D$ non-punishing defectors and $n_{PD}$ punishing defectors, each punishing or non-punishing defector is punished by an amount $(1-\alpha)\rho n_{PC}/(n_{D}+n_{PD}))$).

Individuals gather payoff from playing the game and reproduce with a probability proportional to the exponential of their payoff, such that the population size remains constant. That is, each individual in the next generation is offspring to an individual in the past generation with a probability proportional to the exponential of its payoff. Offspring inherit the strategies of their parent subject to mutations. We assume mutations in the strategies of the individuals to contribute to the public pool and their strategy to contribute to their respective punishing pool occur independently and with the same probability $\nu$. That is, in each reproduction, with probability $\nu$ the offspring of a cooperator (defector) becomes a defector (cooperator), and with the same probability $\nu$, the offspring of a punishing (non-punishing) individual becomes non-punishing (punishing).
\subsection{Structured population}
In the case of a structured population, we assume that individuals live on a network. For most of the simulations, we assume the population network is a first nearest neighbor, two dimensional square lattice with Moore connectivity and periodic boundaries. That is, each site is connected to all the $8$ sites in its vicinity (We will also consider von Neumann connectivity in which each site is connected to four neighboring sites to its right, left, up, and down). Each individual together with its $8$ ‌neighbors form a group. That is a total of $N$ groups in the population. A public goods game followed by a public punishing game is played in each group. That is, each individual belongs to $9$ groups, each centered around himself or one of its $8$ neighbors, and plays $9$ ‌games, one in each group that it belongs to.

The games are played in the same way that was the case for a well-mixed population. That is, in each group, cooperators invest in the public goods game at a cost $c$ to themselves, and defectors do not invest. All the investments in the public pool are multiplied by an enhancement factor $r$, and are divided equally among the group members. In addition, cooperators can invest in a social punishing pool, at a cost $c'$ to themselves. Similarly, defectors can invest to an antisocial punishing pool at a cost $c'$ to themselves. All the investments to the social and antisocial punishing pool are multiplied by a punishment enhancement factor $\rho$, and are used for punishing purposes. As was the case in the well-mixed population, this is done by spending a fraction $\alpha$ of the resources in the social punishing pool to punish cooperators who do not contribute to the social punishing pool, and the rest (a fraction $1-\alpha$ of the resources) is spent to punish defectors. Similarly, a fraction $\alpha$ of the investments in the antisocial punishing pool is spent to punish defectors who do not contribute to the punishing pool, and the rest, a fraction $1-\alpha$, is spent to punish cooperators.

After gathering payoff from the games, individuals reproduce with a probability proportional to the exponential of their payoff. For the reproduction, we consider a synchronous update of the network, in which the whole population is updated at the same time. In addition, we consider a \textit{death-birth} update rule. In this update rule, a new individual at each site is an offspring to an individual living in the extended neighborhood of that site. The extended neighborhood of a site is composed of the focal site, together with its neighboring sites. Mutations can occur as well. Similarly to the case of a well mixed population, we assume mutations in the decisions of the individuals to contribute to the public pool, and their decision to contribute to the punishing pool occur independently, each with probability $\nu$. That is, with probability $\nu$ the offspring of a cooperator (defector) becomes a defector (cooperator), and with probability $\nu$, a punishing (non-punishing) individual become non-punishing (punishing).

We note that, this dynamics can be considered as an imitation with mutation process as well, in which each individual imitates the strategy of one of the individuals in its extended neighborhood, with a probability proportional to the exponential of their payoff, subject to mutations. 

\subsection{The non-wasteful punishment model}
In the model presented in the main text, if no non-punishing cooperators exist in a group, a fraction $\alpha$ of the resources in the social punishing pool is wasted and is not used for other purposes. Similarly, if no defectors exist in a group, a fraction $1-\alpha$ of the resources in the social punishing pool is wasted. The same is true for the resources in the anti-social punishing pool. That is, if no non-punishing defector exist in a group, a fraction $\alpha$ of the resources in the anti-social punishing pool is wasted, and if no cooperators exist in a group, a fraction $1-\alpha$ of the resources in the anti-social punishing pool is wasted. In a more realistic, and at the same time a more complex model, such unused resources could be used for other purposes. One can expect such adaptivity in spending the resources in the punishing pool helps the functioning of the punishing institution, and thus, helps the evolution of cooperation. To see this is indeed the case, here we consider a simple such adaptive or non-wasteful punishment model. In this model, if there is nobody to punish, the resources in the punishment pool are divided among the contributors to the punishment pool. More precisely, in the non-wasteful punishment model, a fraction $\alpha$ of the resources in the social punishment pool is spent to punish non-punishing cooperators. However, if there is no non-punishing cooperator in the group, instead of being wasted, this is divided equally among the contributors to the social punishing pool (i.e. among the punishing cooperators). Similarly, a fraction $1-\alpha$ of the resources in the social punishing pool is spent to punish defectors. However, if there is no defector to punish in a group, this fraction is divided equally among the contributors to the social punishing pool. The same holds for the anti-social punishing pool. That is, a fraction $\alpha$ of the resources in the anti-social punishing pool is spent to punish non-punishing defectors. However, if there is no non-punishing defector in the group, this is divided equally among the contributors to the anti-social punishing pool (i.e. among the punishing cooperators). Similarly, a fraction $1-\alpha$ of the resources in the anti-social punishment pool is spent to punish cooperators. However, if there is no cooperator to punish in a group, this fraction is divided equally among the contributors to the anti-social punishment pool.

All the other details in this model are as before. That is, individuals gather payoff from the game and reproduce with a probability proportional to the exponential of their payoff, such that the population size remains constant. Each individual in the next generation is offspring to an individual in the past generation with a probability proportional to the exponential of its payoff. Offspring inherit the strategies of their parent subject to mutations. Mutations in the decision of the individuals to contribute to the public pool, and their decision to contribute to the punishing pool occur independently, each with probability $\nu$.

\section{Replicator dynamics for the non-wasteful punishment model}
The replicator dynamics for the non-wasteful model can be written using similar argument to that used for the previous model. To do this, we begin by the replicator-mutation equation in the general case:
\begin{align}
m_{x}(t+1)=\sum_{x'}\nu_{x}^{x'}m_{x'}(t)\frac{w_{x'}(t)}{\sum_{x''}m_{x''}(t)w_{x''}(t)}.
\label{eqrepMadaptive}
\end{align}
Here, the same notation as before is used. For completeness we explain the notation here. $x$, $x'$, and $x''$ refer to the strategies, and can be $PC$, $C$, $PD$, or $D$, referring to, respectively, punishing cooperators, non-punishing cooperators, punishing defectors, and non-punishing defectors. $m_x$ is the density of the strategy $x$, $w_{x}$ is the expected fitness of an individual with strategy $x$, and $\nu_{x}^{x'}$ is the mutation rate from the strategy $x'$ to the strategy $x$. Under our assumption that mutations in the strategies of the individuals in the public good pool, and in the public punishing pool occur independently, these can be written in terms of the probability of mutation $\nu$, as follows. For those transformations which require no mutations, that is $x=x'$, we have $\nu_{x}^{x'}=1-2\nu+\nu^2$ (this is the probability that no mutation, neither in the strategy to contribute to the public pool, nor in the strategy to contribute to the punishing pool, occurs). For those  rates which require two mutations, one in the decision of the individuals to contribute to the public pool, and one in their decision to contribute to the punishing pool, we have $\nu_D^{PC}=\nu_C^{PD}=\nu_{PC}^D=\nu_{PD}^C=\nu^2$. All the other rates, which require only one mutation, are equal to $\nu_C^D=\nu_D^C=\nu_{PC}^{PD}=\nu_{PD}^{PC}=\nu_C^{PC}=\nu_{PC}^C=\nu_{PD}^D=\nu_{D}^{PD}=\nu-\nu^2$.

To proceed, we need expressions for the expected fitness of different strategies. These are given by the following equations:
\begin{align}
w_{PC}&=\sum_{n_{PD}=0}^{g-1-n_{PC}-n_{C}}\sum_{n_{C}=0}^{g-1-n_{PC}}\sum_{n_{PC}=0}^{g-1}\exp\bigg[r\frac{1+n_{C}+n_{PC}}{g} -(1-\alpha)\rho\frac{n_{PD}}{1+n_{PC}+n_{C}}+\delta_{n_C,0}\alpha\rho+\delta_{n_{PD}+n_D,0}\nonumber\\&(1-\alpha)\rho-c-c'\bigg]{\rho_{PC}}^{n_{PC}}{\rho_{C}}^{n_{C}}{\rho_{PD}}^{n_{PD}}{\rho_{D}}^{g-1-n_{PC}-n_{C}-n_{PD}}\binom{g-1}{n_{PC},n_{C},n_{PD},g-1-n_{PC}-n_{C}-n_{PD}},\nonumber\\
w_{C}&=\sum_{n_{PD}=0}^{g-1-n_{PC}-n_{C}}\sum_{n_{C}=0}^{g-1-n_{PC}}\sum_{n_{PC}=0}^{g-1}\exp\bigg[r\frac{1+n_{C}+n_{PC}}{g} -(1-\alpha)\rho\frac{n_{PD}}{1+n_{PC}+n_{C}}-\alpha\rho\frac{n_{PC}}{1+n_{C}}-c\bigg]\nonumber\\&{\rho_{PC}}^{n_{PC}}{\rho_{C}}^{n_{C}}{\rho_{PD}}^{n_{PD}}{\rho_{D}}^{g-1-n_{PC}-n_{C}-n_{PD}}\binom{g-1}{n_{PC},n_{C},n_{PD},g-1-n_{PC}-n_{C}-n_{PD}},\nonumber\\
w_{PD}&=\sum_{n_{PD}=0}^{g-1-n_{PC}-n_{C}}\sum_{n_{C}=0}^{g-1-n_{PC}}\sum_{n_{PC}=0}^{g-1}\exp\bigg[r\frac{n_{C}+n_{PC}}{g} -(1-\alpha)\rho\frac{n_{PC}}{1+n_{PD}+n_{D}}+\delta_{n_D,0}\alpha\rho+\delta_{n_{PC}+n_C,0}\nonumber\\&(1-\alpha)\rho-c'\bigg]{\rho_{PC}}^{n_{PC}}{\rho_{C}}^{n_{C}}{\rho_{PD}}^{n_{PD}}{\rho_{D}}^{g-1-n_{PC}-n_{C}-n_{PD}}\binom{g-1}{n_{PC},n_{C},n_{PD},g-1-n_{PC}-n_{C}-n_{PD}},\nonumber\\
w_{D}&=\sum_{n_{PD}=0}^{g-1-n_{PC}-n_{C}}\sum_{n_{C}=0}^{g-1-n_{PC}}\sum_{n_{PC}=0}^{g-1}\exp\bigg[r\frac{n_{C}+n_{PC}}{g} -(1-\alpha)\rho\frac{n_{PC}}{1+n_{PD}+n_{D}}-\alpha\rho\frac{n_{PD}}{1+n_{D}}\bigg]\nonumber\\&{\rho_{PC}}^{n_{PC}}{\rho_{C}}^{n_{C}}{\rho_{PD}}^{n_{PD}}{\rho_{D}}^{g-1-n_{PC}-n_{C}-n_{PD}}\binom{g-1}{n_{PC},n_{C},n_{PD},g-1-n_{PC}-n_{C}-n_{PD}},\nonumber\\
\label{eqpayoffadaptive}
\end{align}
Here, the same notation as before is used, and $\delta_{a,b}$, is the delta function which is equal to $1$ if $a=b$ and it is zero otherwise. These expressions can be derived using similar arguments to those used for the original model. The only difference between these, and the expressions for the original model appeared in eq. (\blue{\eqpayoff}) are the third and fourth terms in the large bracket in the expressions for $w_{PC}$ and $w_{PD}$. For completeness, we explain how these expressions can be derived. We begin by noting that, the term inside the large brackets give the payoff of a focal individual with a given strategy, in a group where there are $n_{PC}$ punishing cooperators, $n_C$ non-punishing cooperators, $n_{PD}$ punishing defectors, and $n_D=g-1-n_{PC}-n_C-n_{PD}$ non-punishing defectors in the group. We begin by explaining how to derive the terms in the large brackets.

In a group with $n_{PC}$ punishing cooperators, $n_C$ non-punishing cooperators, $n_{PD}$ punishing defectors and $n_D=g-1-n_{PC}-n_C-n_{PD}$ non-punishing defectors, $r\frac{1+n_{PC}+n_C}{g}-c$ is the payoff of punishing and non-punishing cooperators and $r\frac{n_{PC}+n_C}{g}$ is the payoff of punishing and non-punishing defectors from the public goods game. These are the first terms in the large bracket in eq. \eqref{eqpayoffadaptive}. A punishing or non-punishing cooperator, who lives in a group with $n_{PC}$ punishing cooperators, $n_{C}$ non-punishing cooperators, $n_{PD}$ punishing defectors and $n_D=g-1-n_{PC}-n_C-n_{PD}$ non-punishing defectors, receives a punishment from the punishing defectors equal to $(1-\alpha)\rho\frac{n_{PD}}{1+n_{PC}+n_{C}}$. This is the second term in the large bracket in the expressions for $w_{PC}$ and $w_C$. In addition, non-punishing cooperators are punished by punishing cooperators in their group, by an amount equal to $\alpha\rho\frac{n_{PC}}{1+n_{C}}$. This is the third term in the large bracket in the expression for $w_C$. In the case of punishing cooperators, if there are no non-punishing cooperator to punish in the group, the fraction $\alpha$ of the resources in the social punishing pool is not used for punishing purposes, and it is instead redistributed among the punishing cooperators. This is the third term in the expression for $w_{PC}$. Similarly, if there are no (punishing and non-punishing) defectors in the group, the fraction $(1-\alpha)$ of the resources in the social punishing pool is not used for punishing purposes and is redistributed among punishing cooperators. This is the fourth term in the expression for $w_{PC}$.

In the same way, a punishing or non punishing defector, who lives in a group with $n_{PC}$ punishing cooperators, $n_{C}$ non-punishing cooperators, $n_{PD}$ punishing defectors, and $n_D$ non-punishing defectors, receives a punishment from the punishing cooperators equal to $(1-\alpha)\rho\frac{n_{PC}}{1+n_{PD}+n_{D}}$. This is the second term in the large bracket in the expressions for $w_{PD}$ and $w_D$. In addition, non-punishing defectors are punished by punishing defectors in their group, by an amount equal to $\alpha\rho\frac{n_{PD}}{1+n_{D}}$. This is the third term in the large bracket in the expression for $w_D$. In the case of punishing defectors, if there are no non-punishing defector to punish in the group, the fraction $\alpha$ of the resources in the anti-social punishing pool is not used for punishing purposes, and it is instead redistributed among the punishing defectors. This is the third term in the expression for $w_{PD}$. Similarly, if there are no (punishing and non-punishing) cooperators in the group, the fraction $(1-\alpha)$ of the resources in the anti-social punishing pool is not used for punishing purposes and is redistributed among punishing defectors. This is the fourth term in the expression for $w_{PD}$.

Finally, as punishing cooperators contribute to both the public pool and the social punishing pool, they pay a cost of $c+c'$. On the other hand, non-punishing cooperators only pay a cost of $c$ to contribute to the public pool. Similarly, punishing defectors, pay a cost of $c'$ to contribute to the antisocial punishing pool. Non-punishing defectors contribute to none of the pools and pay no cost. As individuals reproduce with a probability proportional to the exponential of their payoff, the expected fitness of a strategy can be defined as the expected value of the exponential of their payoff.

To calculate the expected value of fitness, we note that ${\rho_{PC}}^{n_{PC}}{\rho_{C}}^{n_{C}}{\rho_{PD}}^{n_{PD}}{\rho_{D}}^{g-1-n_{PC}-n_{C}-n_{PD}}$ $\binom{g-1}{n_{PC},n_{C},n_{PD},g-1-n_{PC}-n_{C}-n_{PD}}$, is the probability that a focal individual finds itself in a group with $n_{PC}$ punishing cooperators, $n_{C}$ non-punishing cooperators, $n_{PD}$ punishing defectors, and $n_{D}$ non-punishing defectors. Here, $\binom{g-1}{n_{PC},n_{C},n_{PD},g-1-n_{PC}-n_{C}-n_{PD}}=\frac{(g-1)!}{n_{PC}!,n_{C}!,n_{PD}!,(g-1-n_{PC}-n_{C}-n_{PD})!}$ is the multinational coefficient. This is the number of ways that among the $g-1$ group-mates of a focal individual, $n_{PC}$, $n_{C}$, $n_{PD}$, and $g-1-n_{PC}-n_{C}-n_{PD}$ individuals are respectively, punishing cooperators, non-punishing cooperators, punishing defectors, and non-punishing defectors. Summation over all the possible configurations gives the expected fitness of different strategies. Using the expressions in eq. (\ref{eqpayoffadaptive}) for the expected fitness of different strategies in eq. (\ref{eqrepMadaptive}), we have a set of four equations which gives an analytical description of the model, in the limit of infinite population size. 

\begin{figure}[!ht]
	\includegraphics[width=\linewidth, trim = 125 30 65 25, clip,]{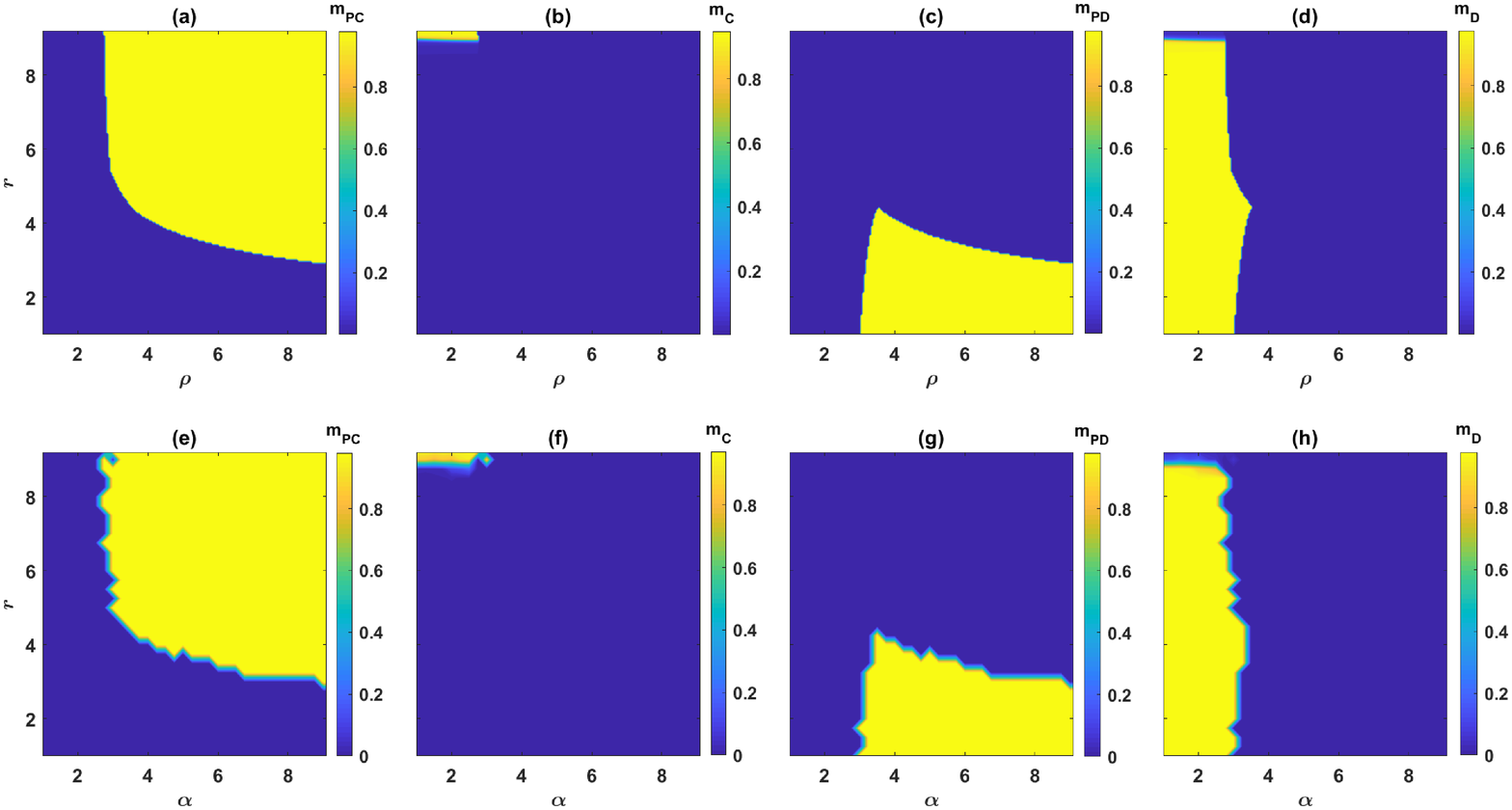}
	\caption{Contour plot of the density of different strategies in a well-mixed population, in the $\rho-r$ plane, resulted from the replicator dynamics ((a) to (d)), and resulted from simulations in a population of size $N=10000$ ((e) to (h)). Here, for the solution of the replicator dynamics, a homogeneous initial condition in which the density of all the strategies equals $0.25$ is used, and for the simulations a random initial condition in which the strategies of the individuals are assigned at random is used. Here, $g=9$, $\nu=10^{-3}$, $c=c'=1$, and $\alpha=0.5$. The simulations are run for $T=200$ time steps, and an average over the last $10$‌ steps of the simulations is taken.}
	\label{figSupr}
\end{figure}
\begin{figure}[!ht]
	\includegraphics[width=\linewidth, trim = 125 30 65 25, clip,]{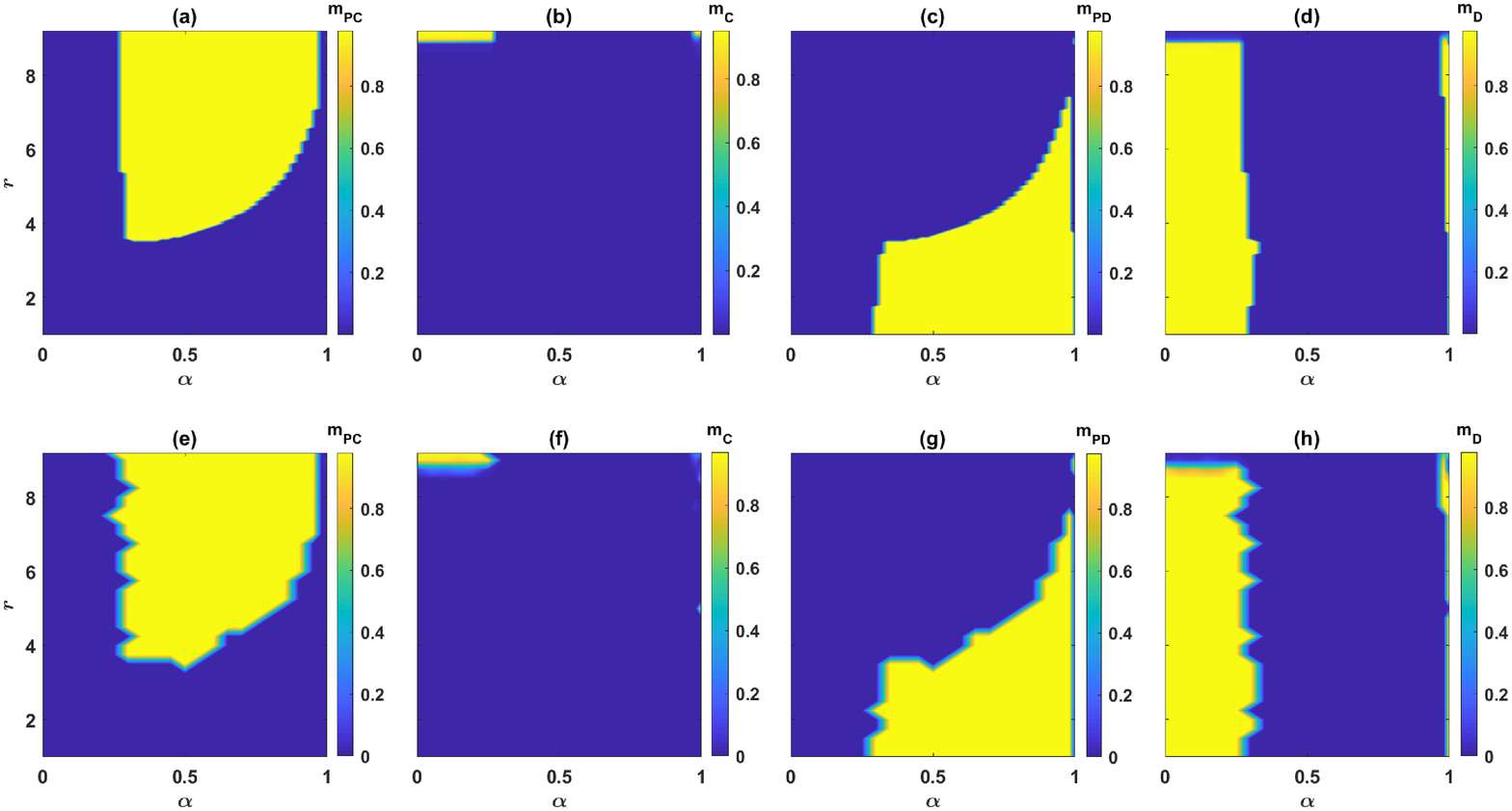}
	\caption{Contour plot of the density of different strategies in a well-mixed population, in the $\alpha-r$ plane, resulted from the replicator dynamics ((a) to (d)), and resulted from simulations in a population of size $N=10000$ ((e) to (h)). Here, for the solution of the replicator dynamics, a homogeneous initial condition in which the density of all the strategies equals $0.25$ is used, and for the simulations a random initial condition in which the strategies of the individuals are assigned at random is used. Here, $g=9$, $\nu=10^{-3}$, $c=c'=1$, and $\rho=5$. The simulations are run for $T=200$ time steps, and an average over the last $10$‌ steps of the simulations is taken.}
	\label{figSupalpha}
\end{figure}
\begin{figure}[!ht]
	\includegraphics[width=\linewidth, trim = 100 15 50 15, clip,]{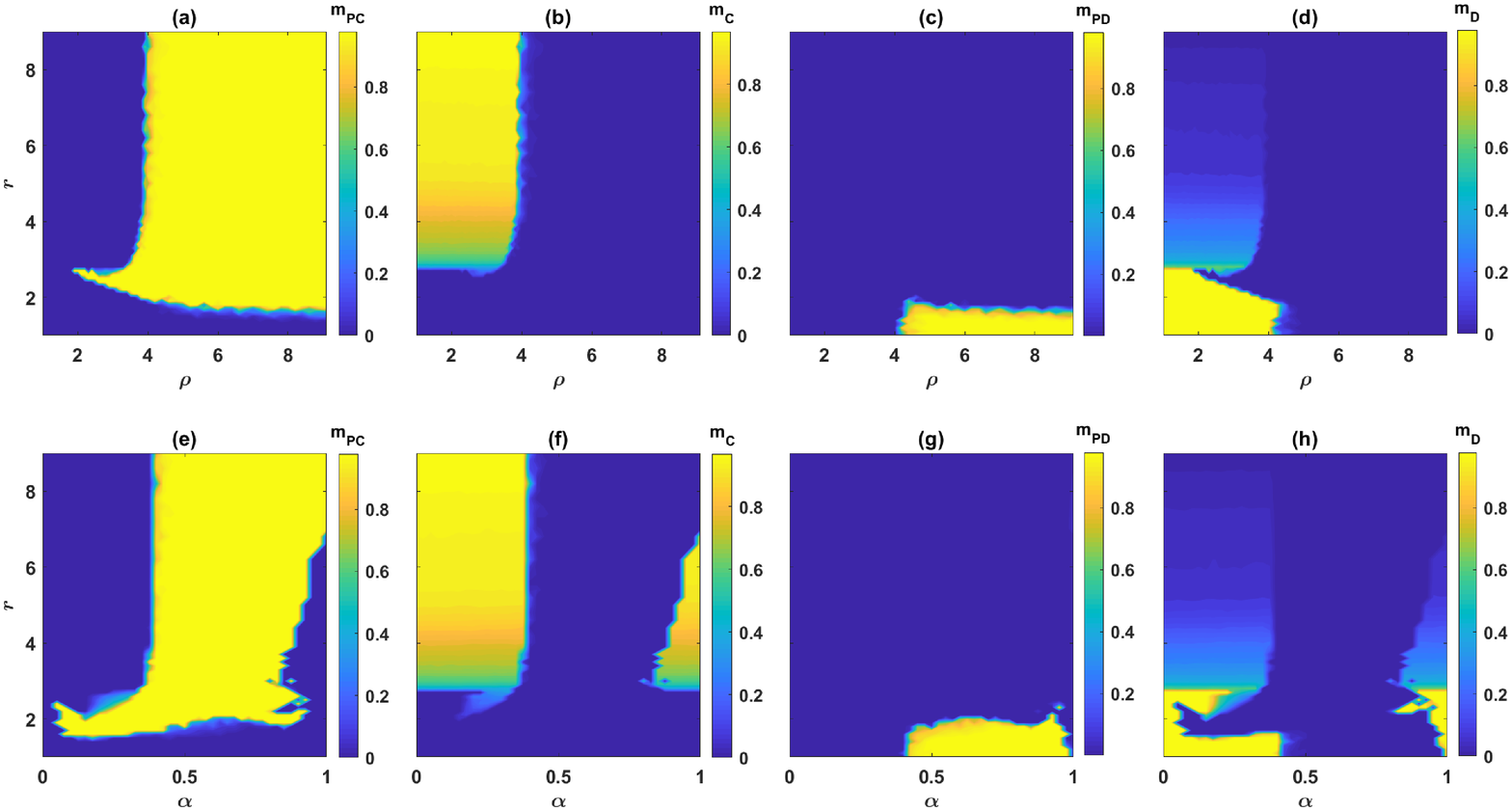}
	\caption{Contour plot of the density of different strategies in a structured population, in the $\rho-r$ plane ((a) to (d)), and in the $\alpha-r$ plane ((e) to (h)). Here, a population of size $N=40000$ individuals residing on a $200\times200$ square lattice with Moore connectivity and periodic boundaries is considered. The simulations are started from a random initial condition in which the strategies of the individuals are assigned at random. Here, $\nu=10^{-3}$, $c=c'=1$, in (a) to (d) $\alpha=0.5$ and in (e) to (h), $\rho=5$. The simulations are run for $T=1000$ time steps, and an average over the last $100$ time steps is taken.}
	\label{figSupnet}
\end{figure}
\section{Analysis of the model: mixed population}
\subsection{The density of different strategies}
The density of different strategies in the $r-\rho$ plane is plotted in Fig. (\ref{figSupr}). In Fig. (\ref{figSupr}.a) to Fig. (\ref{figSupr}.d), numerical solutions of the replicator dynamics are used, and in Fig. (\ref{figSupr}.e) to Fig. (\ref{figSupr}.h), the results of simulations in a population of size $N=10000$ is presented. Fig. (\ref{figSupr}.a) and Fig. (\ref{figSupr}.e) represent the density of punishing cooperators, $m_{PC}$, Fig. (\ref{figSupr}.b) and Fig. (\ref{figSupr}.f) show th density of non-punishing cooperators, $m_C$, Fig. (\ref{figSupr}.c) and Fig. (\ref{figSupr}.g) show the density of punishing defectors, $m_{PD}$, and Fig. (\ref{figSupr}.d) and Fig. (\ref{figSupr}.h) show the density of non-punishing defectors, $m_{D}$. For the solution of the replicator dynamics, a homogeneous initial condition in which the density of all the strategies equals $0.25$ is used, and for the simulations, a random initial condition in which the strategies of the individuals are assigned at random is used. Here, we have set $g=9$, $\nu=10^{-3}$, and $\alpha=0.5$. The phase diagram presented in Fig. (\blue{\figmixed}.a) in the main text are derived from these results by locating the transition lines and different phases.

As can be seen, the results of the replicator dynamics are in high agreement with the result of simulations. Both the replicator dynamics and the simulations show that the system has four different phases. For $r$ smaller than a value close to $g=9$, and for small values of $\rho$, such that the return to the investments in the punishing pool are small, the system goes to a phase where only non-punishing defectors survive. For larger values of $\rho$, such that the return to the investments in the punishing pool is high enough, punishing strategies evolve. In this region, for smaller values of $r$, punishing defectors survive and dominate the population. On the other hand, as $r$ increases, the dynamics settle in a phase where punishing cooperators drive other strategies to extinction and dominate the population. The value of $r$ for which the transition between these two phases occurs depends on $\rho$ ‌and decreases with increasing $\rho$.

The densities of different strategies in the $r-\alpha$ plane are presented in Fig. (\ref{figSupalpha}). Fig. (\ref{figSupalpha}.a) to Fig. (\ref{figSupalpha}.d) represent the numerical solutions of the replicator dynamics, and in Fig. (\ref{figSupalpha}.e) to Fig. (\ref{figSupr}.h), the result of a simulation in a population of size $N=10000$ is presented. Fig. (\ref{figSupalpha}.a) and Fig. (\ref{figSupalpha}.e) represent $m_{PC}$, Fig. (\ref{figSupalpha}.b) and Fig. (\ref{figSupalpha}.f) represent $m_C$, Fig. (\ref{figSupalpha}.c) and Fig. (\ref{figSupalpha}.g) represent $m_{PD}$, and Fig. (\ref{figSupalpha}.d) and Fig. (\ref{figSupalpha}.h) show $m_{D}$. For the solution of the replicator dynamics, a homogeneous initial condition in which the density of all the strategies equals $0.25$ is used, and for the simulations a random initial condition in which the strategies of the individuals are assigned at random is used. Here, as before, we have set $g=9$, $\nu=10^{-3}$, and $\rho=5$. The phase diagrams presented in Fig. (\blue{\figmixed}.b) in the main text, are derived from these results by locating the transition lines and different phases.

We note that for small $\alpha$, such that the punishment of second-order free riders is not sufficiently strong, punishment does not evolve: Non-punishing defectors dominate the population as long as the enhancement factor, $r$, is smaller than a value close to $g=9$. Punishment evolves, as $\alpha$ increases beyond a threshold. In this region, the punishment of second-order free riders is strong enough to overcome the benefit of second-order free riding. However, whether social or anti-social punishment evolves depends on the enhancement factor of the PGG, $r$. For large values of $r$, social punishers dominate the population. On the other hand, for small values of $r$, antisocial punishment evolves. Interestingly, for very large values of $\alpha$, the evolution of social punishment requires a larger value of $r$. This shows, an optimal value of $\alpha$, that is an optimal weight of second-order with respect to first-order punishment exists which facilitates the evolution of social punishment, in the sense that for such an optimal value of $\alpha$, the transition to the social punishment phase occurs for a smaller value of $r$.

Finally, for too large values of $\alpha$, punishing strategies do not evolve, and non-punishing defectors dominate the population. Our analysis thus shows, for social punishment to evolve, it is necessary that both first-order free-riders (those who do not contribute to the public pool), and second-order free-riders (those who do not contribute to the punishing pool) to be punished. Similarly, for antisocial punishment to evolve, it is necessary that both cooperators and non-punishing defectors to be punished.

\subsection{The nature of the phase transitions}
\begin{figure}
	\includegraphics[width=\linewidth, trim = 110 310 105 30, clip,]{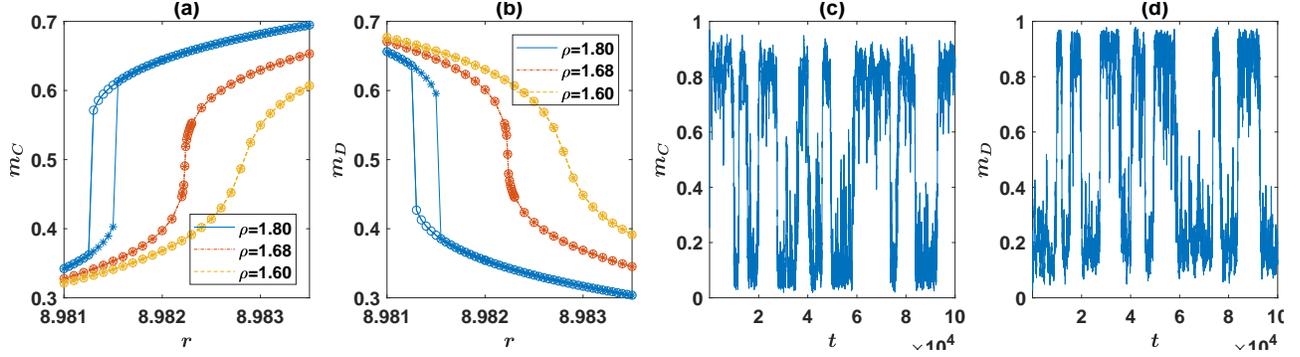}
	\caption{The nature of the $C-D$ transition in a well-mixed population. (a) and (b): the equilibrium density of non-punishing cooperators, $m_C$, (a), and the equilibrium density of non-punishing defectors $m_D$, (b), as a function of $r$ for three different values of $\rho$. Two different initial conditions are used: one in which all the individuals are non-punishing cooperators (circles) and one in which all the individuals are non-punishing defectors (stars). While for large $\rho$, close to the $C-D$ transition, the system shows bi-stability indicating a discontinuous transition, there is a cross-over from the $D$ phase to the $C$ phase for small $\rho$. In between, the transition becomes a continuous transition at a critical value of $\rho$. (c) and (d): The density of non-punishing cooperators, $m_C$, (a), and the density of non-punishing defectors $m_D$, (b), resulted from a simulation in a population of size $N=10000$, as a function of time.
		The dynamics show intermittency between the two phases. In all the cases, $g=9$, $\nu=0.001$, $c=c'=1$, and $\alpha=0.5$. In (c) and (d), $\rho=2$, and $r=9$.}
	\label{figureMixedDCtrns}
\end{figure}

The system is multistable in the whole region of the phase diagram, and depending on the initial conditions, the dynamics settle into one of the stable phases. For $r<g$, three stable phases exist. $PC$ phase, in which punishing cooperators dominate and drive all the other strategies into extinction, $PD$ phase, in which punishing defectors dominate, and the $D$ phase, in which non-punishing defectors dominate. For $r>g$, $PD$ and $D$ phases become unstable. In this region, two stable phases exist, the $C$ phase, in which non-punishing cooperators dominate, and the $PC$ phase. The multi-stability of the system, implies that the $D-PD$ transition, the $C-PC$ transition, and the $PD-PC$ transition, are discontinuous. The $C-D$ transition shows a different behavior compared to the other transitions. While for large $\rho$ this transition is a discontinuous transition, for small $\rho$ there is cross over from the $D$ phase to the $C$ phase, without passing any singularity. In between, there is a critical point where this transition becomes a continuous transition. This can be seen in Fig. (\ref{figureMixedDCtrns}.a) and Fig. (\ref{figureMixedDCtrns}.b), where the density of, respectively, non-punishing cooperators, $m_C$, and non-punishing defectors, $m_D$, as a function of $r$, for three different values of $\rho$, close to the $C-D$ transition are plotted. Here, the replicator dynamics is solved, setting $g=9$ and $\nu=0.001$, and starting from two different initial conditions. As a first initial condition, the initial density of non-punishing cooperators is one, and all the other strategies have a density of zeros in the population. The equilibrium state of this initial condition is indicated in the figure by circles. As the second initial condition, we solve the replicator dynamics starting from an initial condition in which the initial density non-punishing defectors equals one, and all the density of all the other strategies is equal to zero. This is indicated by stars in the figure.

As can be seen, for large $\rho$ ($\rho=1.8$), close to the $C-D$ transition, the system shows bistability. This indicates that the $C-D$ transition is discontinuous in this region.  However, we note that, contrary to the other phases, the coexistence region of the $C$ and the $D$ phases spans a small region close to this transition. On the other hand, for small values of $\rho$ ($\rho=1.6$), as $r$ increases, the system shows a cross over from the $D$ phase to the $C$ phase withought passing any singularity. In between, the transition becomes a singular transition, at a critical value of $\rho$ and $r$.

\begin{figure}%[!ht]
	\includegraphics[width=\linewidth, trim = 110 310 105 30, clip,]{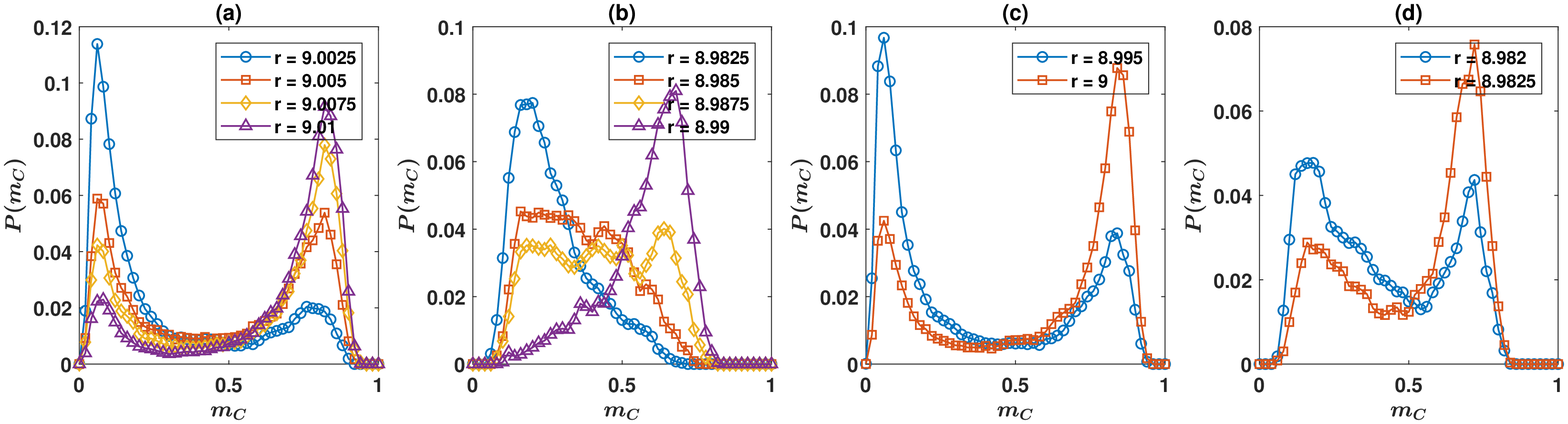}
	\caption{The nature of the $C-D$ transition in a well-mixed population. The distribution of $m_C$ for $\rho=1.25$, (a) and (b), and for $\rho=2$, (c) and (d), resulting from a simulation. In (a) and (c) the population size is $N=10000$ and in (b) and (d) the population size is $N=80000$. While in a small population both distributions show bi-modality, in larger populations, the two cases are different. For small $\rho$, (b), the distribution becomes uni-modal, and for large $\rho$, (d), the distribution remains bi-modal in larger populations, as well. This shows the $C-D$ transition possesses strong finite size effects. Here, $g=9$, $\nu=0.001$, $c=c'=1$, and $\alpha=0.5$. The distribution are derived from a simulation of the system run for $T=150000$ time steps, after discarding the first $1000$ time steps.}
	\label{figureMixedDCTr}
\end{figure}

The results of simulations appear to show a similar phenomenology, as expected. An example of time series of the system close to the $C-D$ transition, resulting from a simulation in a population of size $N=10000$ is presented in Fig. (\ref{figureMixedDCtrns}.c) and Fig. (\ref{figureMixedDCtrns}.d). Here, we plot the density of non-punishing cooperators $m_C$, and the density of non-punishing defectors $m_D$, as a function of time. Here, we have fixed $g=9$, $\rho=2$, and $\nu=0.001$, and have chosen $r=9$, which lies close to the $C-D$ transition. As can be seen, in a finite population, the system shows intermittency between the two phase, which suggest a discontinuous transition. However, simulations in small populations even for small values of $\rho$ show a similar intermittency. This can be seen in Fig. (\ref{figureMixedDCTr}.a), where the distribution of $m_C$, resulted from a simulation in a population of size $N=10000$ is plotted. Here $\rho=1.25$. In Fig. (\ref{figureMixedDCTr}.c), the same distribution for a population of size $N=10000$ for $\rho=2$ is plotted. As can be seen in both cases, the distribution is bi-modal, which should be the case in a discontinuous transition. However, going to larger system sizes, the situation is different for small and large $\rho$. This can be seen in Fig. (\ref{figureMixedDCTr}.b) and Fig. (\ref{figureMixedDCTr}.d), where the distribution of $m_C$ for respectively, $\rho=1.25$ and $\rho=2$, for a population of size $N=80000$ is plotted. As can be seen, going to a large population, while the distribution of $m_C$ remains bi-modal for large $\rho$, it becomes a uni-modal distribution for small $\rho$, as expected from the solutions of the replicator dynamics. We conclude that, close to the $C-D$ transition, the system shows strong finite size effect such the the $C-D$ transition appears to be discontinuous even for small $\rho$, in a small size system.

\section{Analysis of the model: structured populations}
\subsection{The density of different strategies}
The densities of different strategies in the case of the structured population are presented in Fig. (\ref{figSupnet}). In Fig. (\ref{figSupnet}.a) to Fig. (\ref{figSupnet}.d), we present the densities of different strategies in the $r-\rho$ plane, and in Fig. (\ref{figSupnet}.e) to Fig. (\ref{figSupnet}.h), we present the densities of different strategies in the $r-\alpha$ plane. Fig. (\ref{figSupnet}.a) and Fig. (\ref{figSupnet}.e) present $m_{PC}$, Fig. (\ref{figSupnet}.b) and Fig. (\ref{figSupnet}.f) present $m_C$, Fig. (\ref{figSupnet}.c) and Fig. (\ref{figSupnet}.g) present $m_{PD}$, and Fig. (\ref{figSupnet}.d) and Fig. (\ref{figSupnet}.h) show $m_{D}$. Here, simulations are performed on a population of size $N=40000$ individuals residing on a $200\times 200$ first nearest neighbor two dimensional square lattice, with Moore connectivity and periodic boundaries. In Fig. (\ref{figSupnet}.a) to Fig. (\ref{figSupnet}.d), we have set $g=9$, $\nu=10^{-3}$, and $\alpha=0.5$, and in Fig. (\ref{figSupnet}.e) to Fig. (\ref{figSupnet}.h), we have set $g=9$, $\nu=10^{-3}$, and $\rho=5$. The phase diagrams presented in the main text are derived from these results by locating the transition lines and different phases.
\begin{figure}
	\includegraphics[width=\linewidth, trim = 46 185 56 15, clip,]{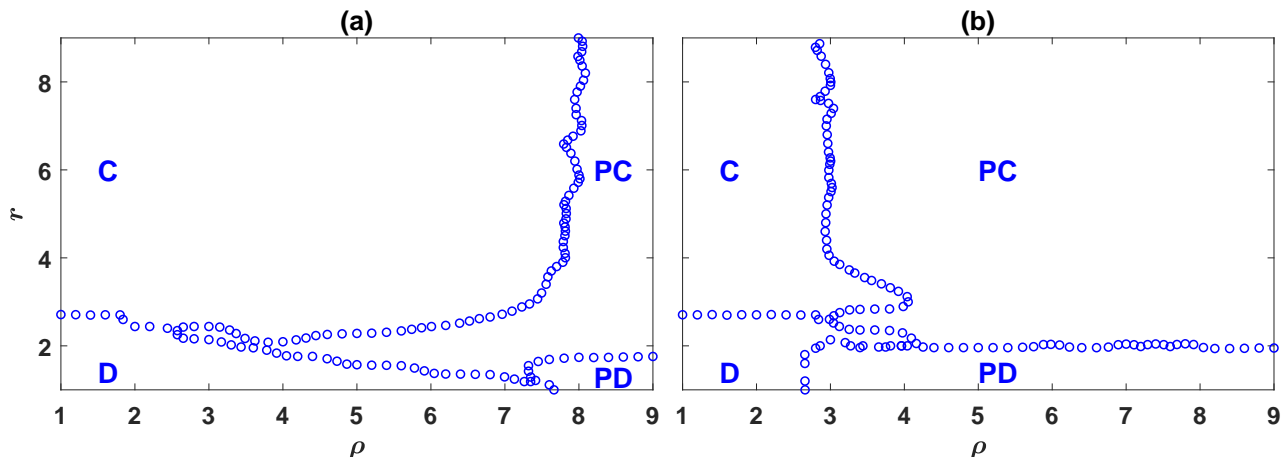}
	\caption{Evolution of social punishment is facilitated near a phase transition. The phase diagram of the model for a structured population for two different values of $\alpha$. In (a), $\alpha=0.25$ and in (b), $\alpha=0.75$. For both values of $\alpha$, close to the $C-D$ phase transition, the transition to the social punishment requires smaller punishment enhancement factors. The phase diagram is derived by running simulations in a population of size $40000$, residing on a $200 \times 200$ first nearest neighbor lattice with Moore connectivity and periodic boundaries. Here, $g=9$, $\nu=0.001$, and $c=c'=1$.}
	\label{figphasenetCD}
\end{figure}

We begin our analysis, by investigating the behavior of the system in the $r-\rho$ plane. As seen in the phase diagram of the model presented in the main text, the system can settle in four different phases, in each of which one of the four possible strategies dominates. For small $\rho$, such that the return to the investment in the punishment pool is small, non-punishing strategies evolve. Here, for small $r$ the system settles in the $D$ phase, in which non-punishing defectors dominate. By increasing $r$, for a small $\rho$, a phase transition to a phase in which non-punishing cooperators survive (which we call the $C$ phase) occurs. We note that, for a mixed population, this transition occurs for a value of $r$ close to $g=9$, where, investment to the public pool yields a positive payoff to an individual, and thus, it is no longer a dilemma. On the other hand, as can be seen in the figure, for a structured population this transition occurs for a much smaller value of $r$ (smaller than $3$), where investment to the public pool results in a negative payoff for the investor, and thus, constitutes a social dilemma. This shift in the transition is due to the facilitating effect of network structure for the evolution of cooperation. In addition, comparison with a similar model, in which only non-punishing strategies exist, shows the $C-D$ transition shifts to smaller values compared to such a model, with the same size and network structure. This is due to the positive effect of the existence of the punishing strategies for the evolution of cooperation. 

As the value of $\rho$ increases, punishing strategies evolve. Here, for very small $r$, the $PD$ phase, in which punishing defectors dominate the population occurs. As $r$ increases, for fixed large enough $\rho$, a transition to the $PC$ phase in which punishing cooperators dominate the population occurs. We note that, for a fixed large enough $\rho$, the value of $r$ where the transition from the anticosial punishment to the social punishment evolves is smaller than the value or $r$ where the transition from the $D$ phase to the $C$ phase occurs. Furthermore, by increasing $\rho$, the $PD-PC$ transition line shifts to smaller values of $r$. This interesting observation results from the beneficial effect of punishment for the evolution of cooperation, such that the larger $\rho$ and the more effective the punishment, the easier cooperative strategies can evolve.

\begin{figure}%[!ht]
	\includegraphics[width=\linewidth, trim = 100 254 90 20, clip,]{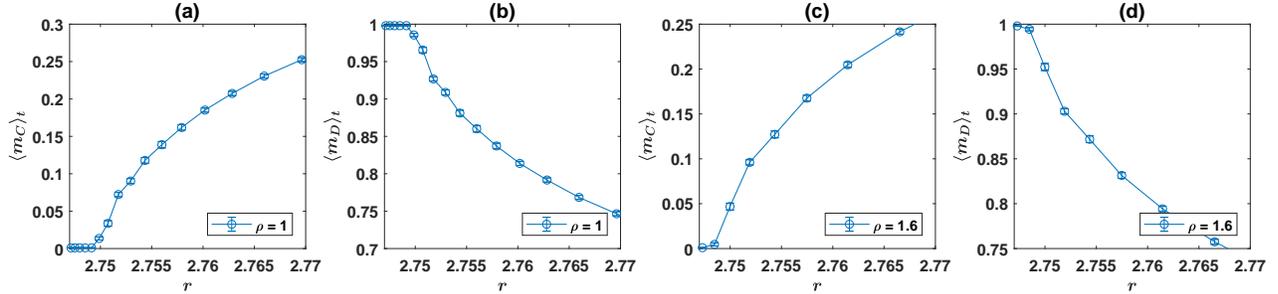}
	\caption{The nature of the $C-D$ transition in a structured population. (a) to (d): The time average densities of non-punishing cooperators (a and c) and non-punishing defectors (b and d) as a function of $r$, close to the $C-D$ transition for two different values of $\rho$ as indicated in the figure. For a fixed small value of $\rho$, for $r$ smaller than a critical value, non-punishing defectors dominate the population. As $r$ increases beyond a critical value, non-punishing cooperators can survive. The transition between the two phases shows no discontinuity and appears continuously, for all values of $\rho$. Here, the simulations are performed in a population of size $N=640000$, living on a $800\times800$ square lattice, with periodic boundaries and Moore connectivity. The mutation rate is set equal to $\nu=0.001$. The simulations are run for $T=50000$ time steps and the time averages are taken over the last $20000$ time steps of the simulations.}
	\label{figDCtrans}
\end{figure}
Finally, we note that the evolution of social punishment is facilitated close to the $C-D$ transition. This can be seen by noting that the value of $\rho$ for which the system settles in the $PC$ phase significantly decreases close to the $C-D$ transition. This interesting observation shows that being close to a physical phase transition, facilitates the evolution of social punishment.

We turn to the densities of different strategies in the $r-\alpha$ plane, presented in Fig. (\ref{figSupnet}.e) to Fig. (\ref{figSupnet}.h). As can be seen in the figure, for small values of $\alpha$, such that the punishment of second-order free-riding is not strong enough, punishing strategies do not evolve. Instead, here for small values of $r$, non-punishing defectors dominate, and as $r$ increases, the system shows a transition to a phase where non-punishing cooperators can survive. Similarly, for small enough values of $r$, and for too large values of $\alpha$, such that the punishment of defectors by social punishers (or the punishment of cooperators by antisocial punishers) is too weak, punishing strategies do not evolve and the system settles in either the $D$ phase (for smaller $r$) or the $C$ phase (for larger $r$). However, when $r$ is very large (larger than $\sim 6$), social punishment can evolve, even when $\alpha=1$. That is when social punishers do not punish defectors. This was not the case in the case of a mixed population and results from strong network reciprocity for very large values of $r$. For values of $\alpha$, in between these two extremes, punishing strategies evolve. Here, for small values of $r$ antisocial punishment evolves, and as $r$ increases, the system shows a phase transition to a phase where social punishment evolves.

\begin{figure}[!ht]
	\includegraphics[width=\linewidth, trim = 110 315 90 30, clip,]{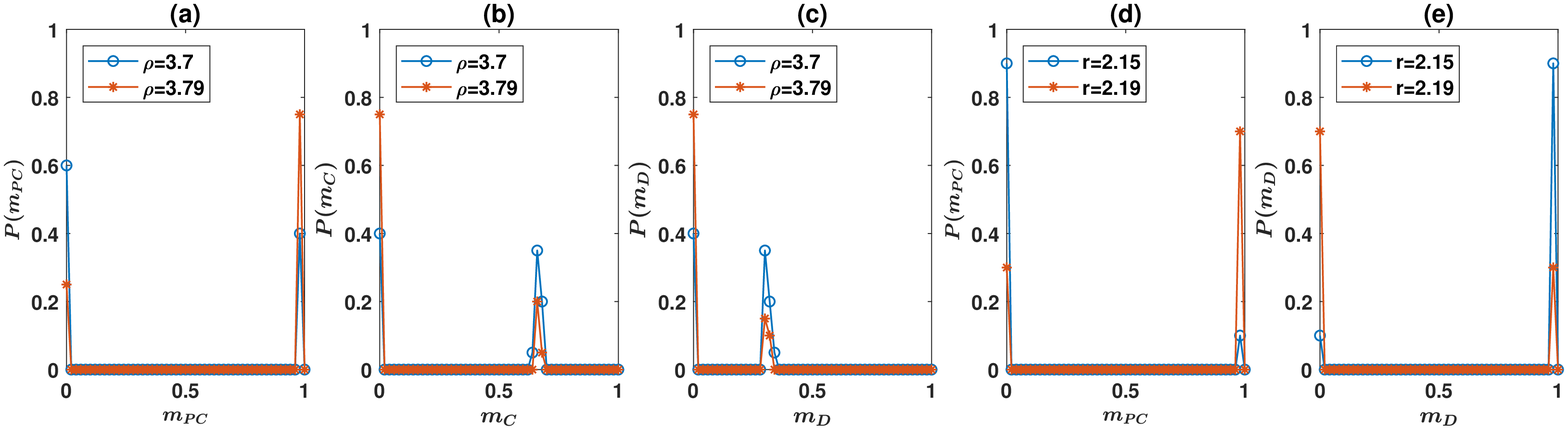}
	\caption{The $C-PC$ and $D-PC$ transitions are discontinuous. (a) to (c): The $C-PC$ transition shows bi-stability and is discontinuous. The distribution of $m_{PC}$ (a), $m_C$ (b), and $m_D$ (c), for $r=4$ and two different values of $\rho$ chosen close to the $C-PC$ transition. All the distributions are bimodal, which shows the $C-PC$ transition is discontinuous. The peak corresponding to a small value of $m_{PC}$, and large values of $m_C$ and $m_D$ corresponds to the $C$ phase, and the peak corresponding to a value of $m_{PC}$ close to $1$, and $m_C$ and $m_D$ close to $0$, corresponds to the $PC$ phase. By increasing $\rho$, the peak corresponding to the $C$ phase decreases, while that corresponding to the $PC$ phase increases. (d) and (e): The distribution of $m_{PC}$ (d) and $m_D$ (e), for $\rho=3.2$ and two different values of $r$, chosen close to the $D-PC$ transition. The distributions are bimodal, which shows the $D-PC$ transition is discontinuous. The peak with $m_{PC}$ close to $0$ and $m_D$ close to $1$ corresponds to the $D$ phase, while that with $m_{PC}$ close to $1$ and $m_D$ close to $0$ corresponds to the $PC$ phase. By increasing $r$ the peak corresponding to the $D$ phase decreases, while that corresponding to the $PC$ phase increases. This shows the $D-PC$ transition is discontinuous. In all the panels, the network is a  two dimensional $100\times100$ square lattice with Moore connectivity and periodic boundaries, and $\nu=0.005$. The distributions are derived from the final state of a sample of $R=20$ simulations, starting from random initial conditions. In (a) to (c), the simulations are run for $T=6000$ time steps, and in (d) and (e), the simulations are run for $T=10000$ time steps.}
	\label{figureCPCandDPC}
\end{figure}
\begin{figure}[!ht]
	\includegraphics[width=\linewidth, trim = 109 315 90 30, clip,]{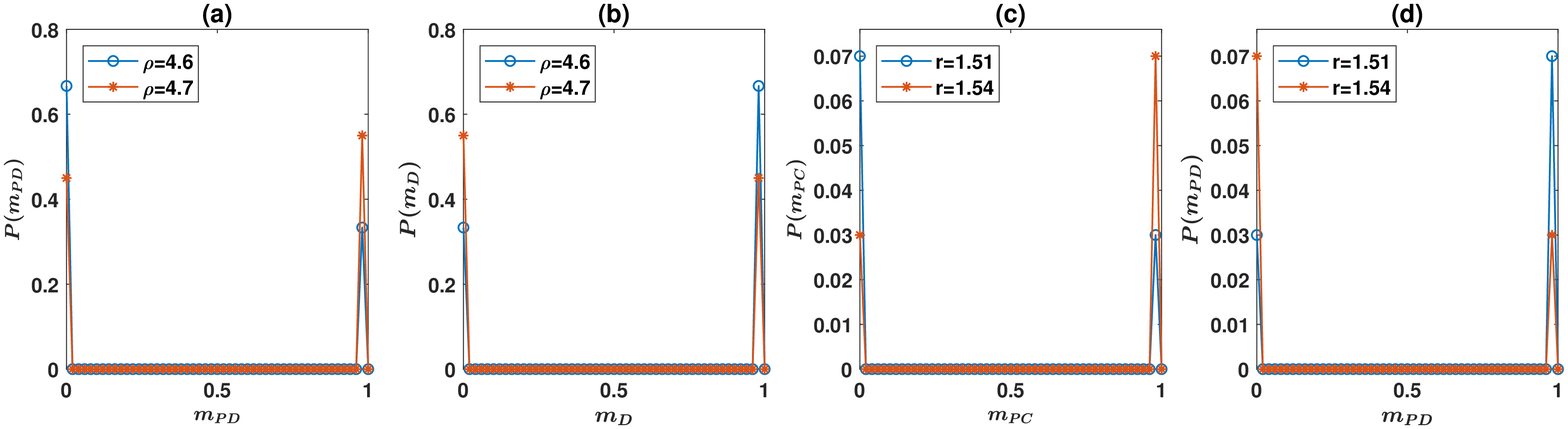}
	\caption{The $D-PD$ and $PD-PC$ transitions are discontinuous. (a) and (b): The distribution of $m_{PD}$ (a), and $m_D$ (b), for $r=1.2$ and two different values of $\rho$ chosen close to the $D-PD$ transition. As can be seen, these distributions are bimodal. The peak corresponding to a value of $m_{PD}$ close to $0$, and $m_D$ close to $1$, corresponds to the $D$ phase, and the peak corresponding to a value of $m_{PD}$ close to $1$, and $m_D$ close to $0$, corresponds to the $PD$ phase. By increasing $\rho$, the peak corresponding to the $D$ phase decreases, while that corresponding to the $PD$ phase increases. This shows the $D-PD$ transition is discontinuous. (c) and (d): The distribution of $m_{PC}$ (c), and $m_{PD}$ (d), for $\rho=5$ and two different values of $r$, chosen close to the $PD-PC$ transition. The distributions are bimodal, which shows the $PD-PC$ transition is discontinuous. The peak with $m_{PC}$ close to $0$ and $m_{PD}$ close to $1$ corresponds to the $PD$ phase, while that with $m_{PC}$ close to $1$ and $m_{PD}$ close to $0$ corresponds to the $PC$ phase. By increasing $r$ the peak corresponding to the $PD$ phase decreases, while that corresponding to the $PC$ phase increases. In all the panels, the network is a two dimensional $100\time100$ lattice with periodic boundaries and Moore connectivity (in (a) and (b)), or Von Neuman connectivity (in (c) and (d)). In (a) and (b) $\nu=0.005$, and in (c) and (d), $\nu=0.001$. The distributions are derived from the final state of a sample of $R=10$ simulations, starting from random initial conditions. In (a) and (b), the simulations for $\rho=4.6$ (blue curve marked with circle) is run for $T=50000$ time steps, and that for $\rho=4.7$ (red curve marked with stars) is run for $T=75000$ time steps, and in (d) and (e), the simulations are run for $T=10000$ time steps. In (c) and (d), the simulations are run for $T=30000$ time steps.}
	\label{figureDPDandPDPC}
\end{figure}

\subsection{The evolution of social punishment is facilitated close to a phase transition}
We have seen that close to the $C-D$ phase transition, the evolution of social punishment is facilitated, as it happens for smaller values of the punishment enhancement factor, $\rho$. In this section we argue that this result is robust and occurs for different parameter values. To do this, we run simulations for two different values of $\alpha$ to drive the phase diagram of the system. The results are presented in Fig. (\ref{figphasenetCD}). In Fig. (\ref{figphasenetCD}.a), $\alpha=0.25$ and in Fig. (\ref{figphasenetCD}.b), $\alpha=0.75$. Here, the phase diagram is derived by running simulations in a population of size $40000$, residing on a $200 \times 200$ first nearest neighbor lattice with Moore connectivity and periodic boundaries. Here, $g=9$, and $\nu=0.001$. As can be seen, for both values of $\alpha$, close to the $C-D$ phase transition, the transition to the social punishment requires smaller punishment enhancement factors. The shift is more significant for smaller $\alpha$, where away from the phase transition, a relatively large value of $\rho$ is necessary for the evolution of both social and antisocial punishment. In contrast, the transition to the social punishment phase shifts to significantly smaller values of $\rho$ close to the $C-D$ phase transition.

For larger values of $\alpha$, as in Fig. (\ref{figphasenetCD}.b), the evolution of social punishment is facilitated and it can happen for smaller values of $\rho$. In addition, for larger enhancement factors, $r$, the transition to the social punishment is easier as it requires smaller punishment enhancement factors, $\rho$. However, close to the $C-D$ transition, this transition shifts to smaller values of $\rho$. This suggests that being close to a phase transition facilitates the evolution of social punishment in this case, as well.

\subsection{The nature of the phase transitions}
In this section, we show that all the transitions which involve punishing strategies, that is $C-PC$, $D-PC$, $D-PD$, and $PD-PC$ show bi-stability and are discontinuous. On the other hand, the $C-D$ transition appears to be a continuous transition.

We begin by the $C-D$ transition, which has a different nature from the other transitions. As mentioned before, for each value of $\rho$, small enough, as $r$ increases a phase transition from the $D$ phase, in which non-punishing defectors dominate the population, to the $C$ phase, in which non-punishing cooperators dominate occurs. We call this transition the $C-D$ transition. To study this transition, we perform simulations in a population of size $N=640000$, living on a first nearest neighbor two dimensional $800\times800$ square lattice, with periodic boundaries and Moore connectivity. The results of simulations for two different values of $\rho$ are presented in Fig. (\ref{figDCtrans}). Here, the mutation rate is set $\nu=0.001$, and the densities of non-punishing cooperators and non-punishing defectors are plotted.

As can be seen, for small $r$ non-punishing defectors dominate the population, with a density close to $1$. In this phase, in addition to non-punishing defectors, a small fraction of other strategies are maintained in the population, due to mutations. As $r$ increases beyond a pseudo-critical point, non-punishing cooperators start to survive. As shown below, they do so, by forming compact domains in which the benefit of cooperation is reaped by the fellow cooperators. Close to the pseudo-critical point, by increasing $r$, the density of non-punishing cooperators rapidly increases, and that of the non-punishing defectors rapidly decreases. The change in the densities of all the strategies occurs continuously, and no discontinuity in this transition is observed. This suggests that the $D-C$ transition is a continuous transition.

We begin the study of the discontinuous transitions by the $C-PC$ transition. To this goal, in Fig. (\ref{figureCPCandDPC}.a) to Fig. (\ref{figureCPCandDPC}.c), we plot the distribution of, respectively, $m_{PC}$, $m_C$, and $m_D$. Here, a population of $N=10000$ individuals living on a first nearest neighbor two dimensional $100\times 100$ lattice with Moore connectivity and periodic boundaries is considered. The mutation rate, is set equal to $\nu=0.005$. The distributions are derived from the final state of a sample of $R=20$ simulations, after $T=6000$ time steps of evolution, starting from random initial conditions. Here, we have set $r=4$ and the distributions are plotted for two different values of $\rho$ chosen close to the $C-PC$ transition. As can be seen in the figure, the distributions are bimodal and have two peaks. This shows that the $C-PC$ transition is discontinuous. The peak corresponding to a small value of $m_{PC}$, and large values of $m_C$ and $m_D$ corresponds to the $C$ phase, in which cooperators survive and can coexist with defectors. On the other hand, the peak corresponding to a value of $m_{PC}$ close to $1$, and $m_C$ and $m_D$ close to $0$, corresponds to the $PC$ phase, in which punishing cooperators dominate the population. By increasing $\rho$, the peak corresponding to the $C$ phase decreases, while that corresponding to the $PC$ phase increases. This phenomenology is characteristic of a discontinuous transition.

The nature of the $D-PC$ transition is investigated in Fig. (\ref{figureCPCandDPC}.d), where the distribution of $m_{PC}$ is plotted, and Fig. (\ref{figureCPCandDPC}.e), where the distribution of $m_{D}$ is plotted. Here, a population of $N=10000$ individuals living on a first nearest neighbor two dimensional $100\times 100$ lattice with Moore connectivity and periodic boundaries is considered. The mutation rate is set equal to $\nu=0.005$. The distributions are derived from the final state of a sample of $R=20$ simulations, after $T=10000$ time steps of evolution starting from random initial conditions. Here, we have set $\rho=3.2$ and the distributions are plotted for two different values of $r$, as indicated in the figures. These values are chosen such that the system is posed close to the $D-PC$ transition. As can be seen, the distributions have two peaks and are bimodal. The peak with $m_{PC}$ close to $0$ and $m_D$ close to $1$ corresponds to the $D$ phase, while that with $m_{PC}$ close to $1$ and $m_D$ close to $0$ corresponds to the $PC$ phase. By increasing $r$ the peak corresponding to the $D$ phase decreases, while that corresponding to the $PC$ phase increases. This phenomenology is characteristic of a discontinuous phase transition and shows that the $D-PC$ transition is discontinuous.

The nature of the $D-PD$ transition is investigated in Fig. (\ref{figureDPDandPDPC}.a) and Fig. (\ref{figureDPDandPDPC}.b). In Fig. (\ref{figureDPDandPDPC}.a), the distribution of $m_{PD}$ is plotted, and in Fig. (\ref{figureDPDandPDPC}.b), the distribution of $m_{D}$ is plotted. Here, a population of $N=10000$ individuals living on a first nearest neighbor two dimensional $100 \times 100$ lattice with Moore connectivity and periodic boundaries is considered. The mutation rate, is set equal to $\nu=0.005$. The distributions are derived from the final state of a sample of $R=20$ simulations, starting from random initial conditions. Here, we have set $r=1.2$ and the distributions are plotted for two different values of $\rho$, as indicated in the figures, chosen such that the system is posed close to the $D-PD$ transition. The simulations for $\rho=4.6$ (blue curve marked with circle) is run for $T=50000$ time steps, and that for $\rho=4.7$ (red curve marked with stars) is run for $T=75000$ time steps. As can be seen, the distributions have two peaks and are bimodal. The peak with $m_{PD}$ close to $0$ and $m_D$ close to $1$ corresponds to the $D$ phase, while that with $m_{PD}$ close to $1$ and $m_D$ close to $0$ corresponds to the $PD$ phase. By increasing $\rho$, the peak corresponding to the $D$ phase decreases, while that corresponding to the $PD$ phase increases. This phenomenology is characteristic of a discontinuous phase transition and shows that the $D-PD$ transition is discontinuous.
\begin{figure}
	\includegraphics[width=\linewidth, trim = 50 135 55 20, clip,]{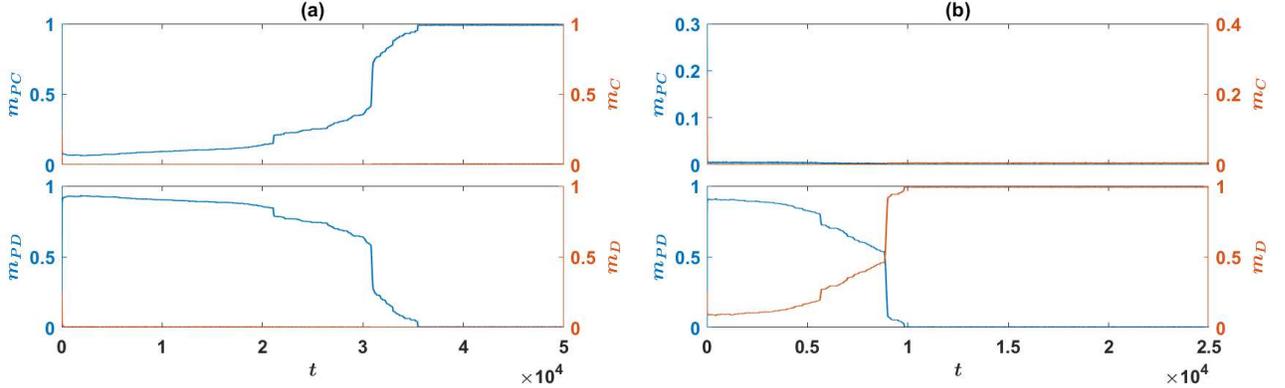}
	\caption{Time evolution of the density of different strategies close to the $D-PC$ transition. In (a), $r=1.72$ and $\rho=5$, and in (b), $r=1.2$ and $\rho=4.65$. In both cases $c=c'=1$, $\nu=0.005$, and the population resides on a $100\times100$ square lattice with Moore connectivity and periodic boundaries. In both cases, the time evolution of the system shows two different time scales: a slow growth, intermittent with short periods of fast growth. This time evolution is characteristic of the system close to all the discontinuous transitions. The slow growth corresponds to slow growth of homogeneous blocks along their vertical and horizontal boundaries, and the fast growth sets in when different blocks merge, which results in non- horizontal and non-vertical boundaries, along which the growth of a homogeneous block happens in a much faster speed.}
	\label{figtime}
\end{figure}

Finally, the nature of the $PD-PC$ transition is investigated in Fig. (\ref{figureDPDandPDPC}.c) and Fig. (\ref{figureDPDandPDPC}.d). In Fig. (\ref{figureDPDandPDPC}.c), the distribution of $m_{PC}$ is plotted, and in Fig. (\ref{figureDPDandPDPC}.d), the distribution of $m_{PD}$ is plotted. Here, a population of $N=10000$ individuals living on a first nearest neighbor two dimensional $100\times100$ lattice with von Neumann connectivity (that is each site is connected to four sites, to its north, south, east, and west) and periodic boundaries is considered. The mutation rate, is set equal to $\nu=0.001$. The distributions are derived from the final state of a sample of $R=10$ simulations, starting from random initial conditions. The simulations are run for $T=30000$ time steps. Here, we have set $\rho=5$ and the distributions are plotted for two different values of $r$, as indicated in the figures, chosen such that the system is posed close to the $PD-PC$ transition. As can be seen in the figure, the distributions are bimodal, which shows the $PD-PC$ transition is discontinuous. The peak with $m_{PC}$ close to $0$ and $m_{PD}$ close to $1$ corresponds to the $PD$ phase, in which punishing defectors dominate. On the other hand, the peak with $m_{PC}$ close to $1$ and $m_{PD}$ close to $0$ corresponds to the $PC$ phase, in which punishing cooperators dominate the population. By increasing $r$ the peak corresponding to the $PD$ phase decreases, while that corresponding to the $PC$ phase increases. This shows the $PD-PC$ transition is discontinuous.

We note that, contrary to the other cases, where we have used a lattice with Moore connectivity, here, a different connectivity, von Neumann connectivity is used. The reason is that, as it will be explained shortly, the dynamics of the system close to the discontinuous transitions is driven by slow growth of domains of individuals with the same strategy, in a sea of individuals with another, competing strategy. On a network with Moore connectivity, these domains are rectangular domains, and the growth of the domains proceeds along their horizontal and vertical boundaries. As close to the transition, such a growth can be very slow, the time needed for the system to reaches equilibrium can be excessively large. This is the case for all the discontinuous transitions, but particularly, prominent for the case of the $PC-PD$ transition. As the nature of the transition is the same for both Moore and von Neumann connectivity, we have changed the network connectivity to von Neumann connectivity, where the equilibration time of the system is shorter, and thus, it is more convenient to study.

\begin{figure}[!ht]
	\includegraphics[width=\linewidth, trim = 115 35 100 30, clip,]{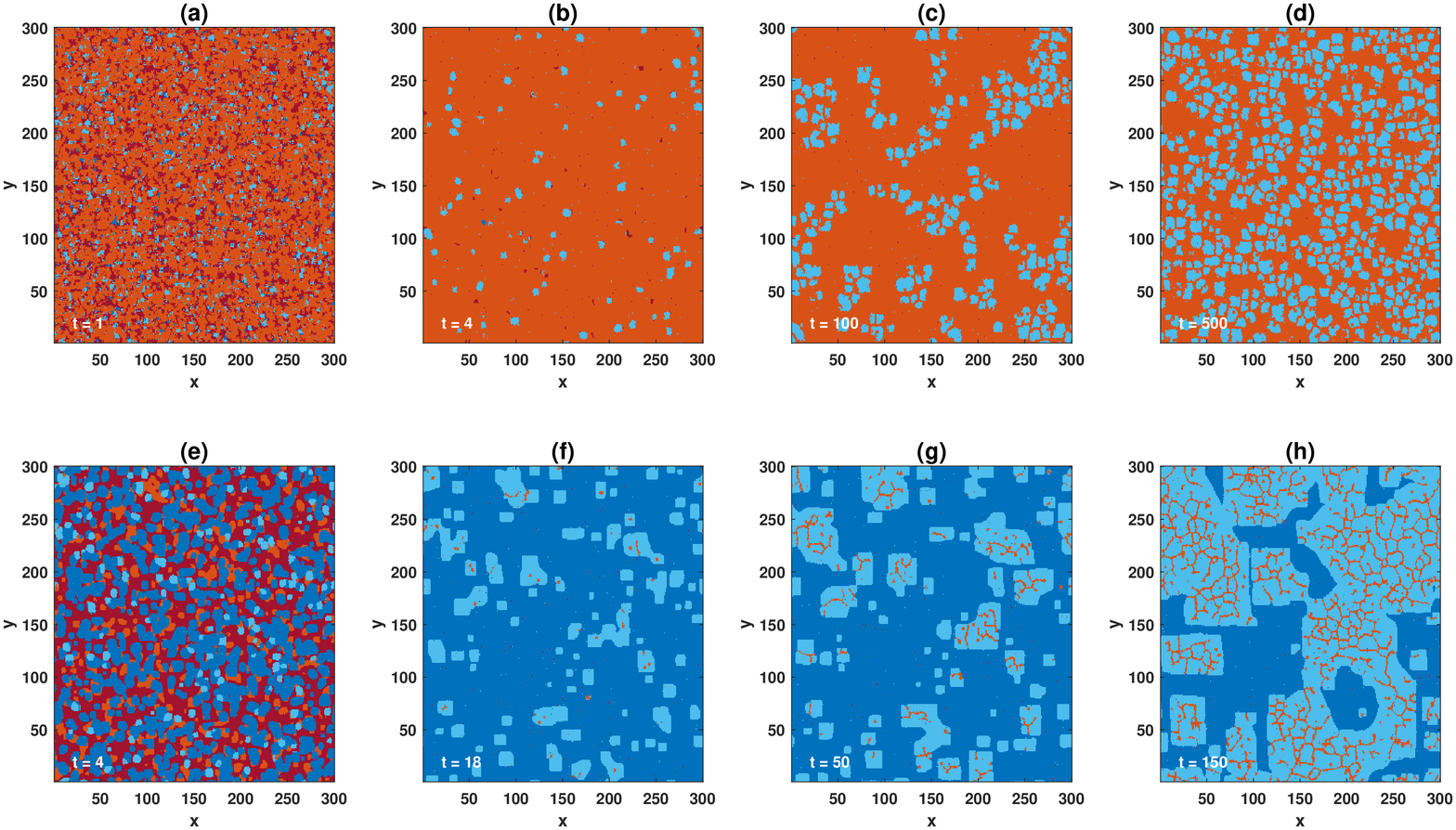}
	\caption{Time evolution of the system, close to the $C-D$ phase transition ((a) to (d)), and close to the $C-CP$ phase transition ((e) to (h)). Different strategies are indicated by different colors. Light blue shows non-punishing cooperators, dark blue shows punishing cooperators, light red shows non-punishing defectors, and dark red shows punishing defectors. (a) to (d): Starting from a random initial condition, close to the $C-D$ transition, defectors rapidly increase in density and dominate the population. While all the other strategies go to extinction, non-punishing cooperators can survive by forming compact domains in which the benefit of cooperation is reaped by fellow cooperators. The dynamics is governed by formation, division, and collapse of cooperator blocks in the sea of defectors. (e) to (h): Close to the $C-PC$ transition, starting from a random initial condition, punishing cooperators rapidly expand by driving all the other strategies into extinction. Only non-punishing cooperators can survive by forming small rectangular domains. These blocks slowly increase in the see of punishing cooperators, until they expand the whole population. While defectors can not survive in the sea of punishing cooperators, they survive by forming narrow bands in the non-punishing cooperators' communities. Here, a population of $N=90000$ individuals lives on a  two dimensional $300\times300$ lattice with periodic boundaries and Moore connectivity. In (a) to (d) $r=2.8$ and $\rho=1.4$, and in (e) to (h), $r=4$ and $\rho=3.47$. In all the cases $\nu=10^{-3}$, and $c=c'=1$.}
	\label{figvideoCDandCCP}
\end{figure}
\begin{figure}[!ht]
	\includegraphics[width=\linewidth, trim = 115 35 100 30, clip,]{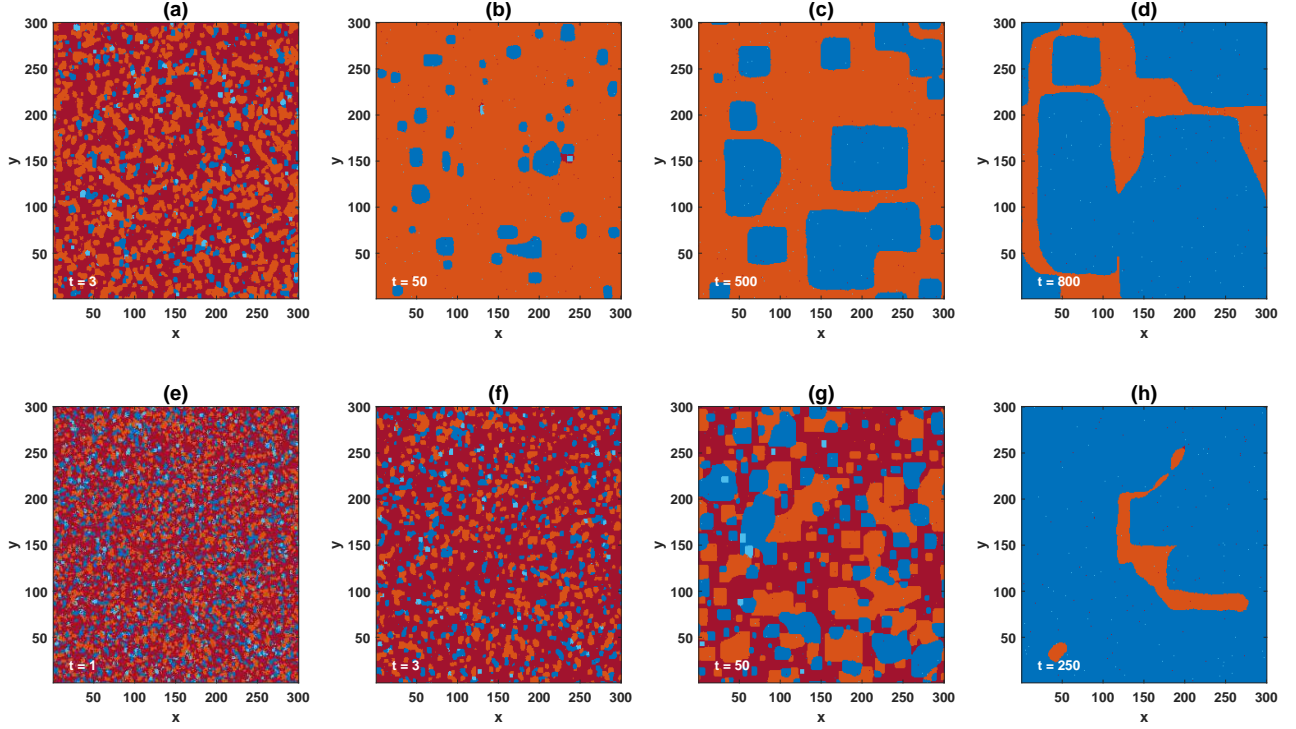}
	\caption{Time evolution of the system, close to the $D-PC$ phase transition. In (a) to (d), $r=2.2$ and $\rho=3.2$, and in (e) to (h), $r=2.1$ and $\rho=3.8$. Different strategies are indicated by different colors. Light blue shows non-punishing cooperators, dark blue shows punishing cooperators, light red shows non-punishing defectors, and dark red shows punishing defectors. Starting from a random initial condition, close to the $C-PC$ transition, one of the strategies (non-punishing defectors in (a) to (d), and punishing defectors in (e) to (h)) rapidly increase in density and form a sea of homogeneous strategies. Other strategies can only survive by forming compact domains. The second stage of the evolution is governed by slow growth of domains of punishing cooperators along their boundaries. In both cases, a population of $N=90000$ individuals lives on a two dimensional $300\times300$ lattice with periodic boundaries and Moore connectivity. Here, $\nu=10^{-3}$, and $c=c'=1$.}
	\label{figvideoDPC}
\end{figure}
\begin{figure}[!ht]
	\includegraphics[width=\linewidth, trim = 115 35 100 30, clip,]{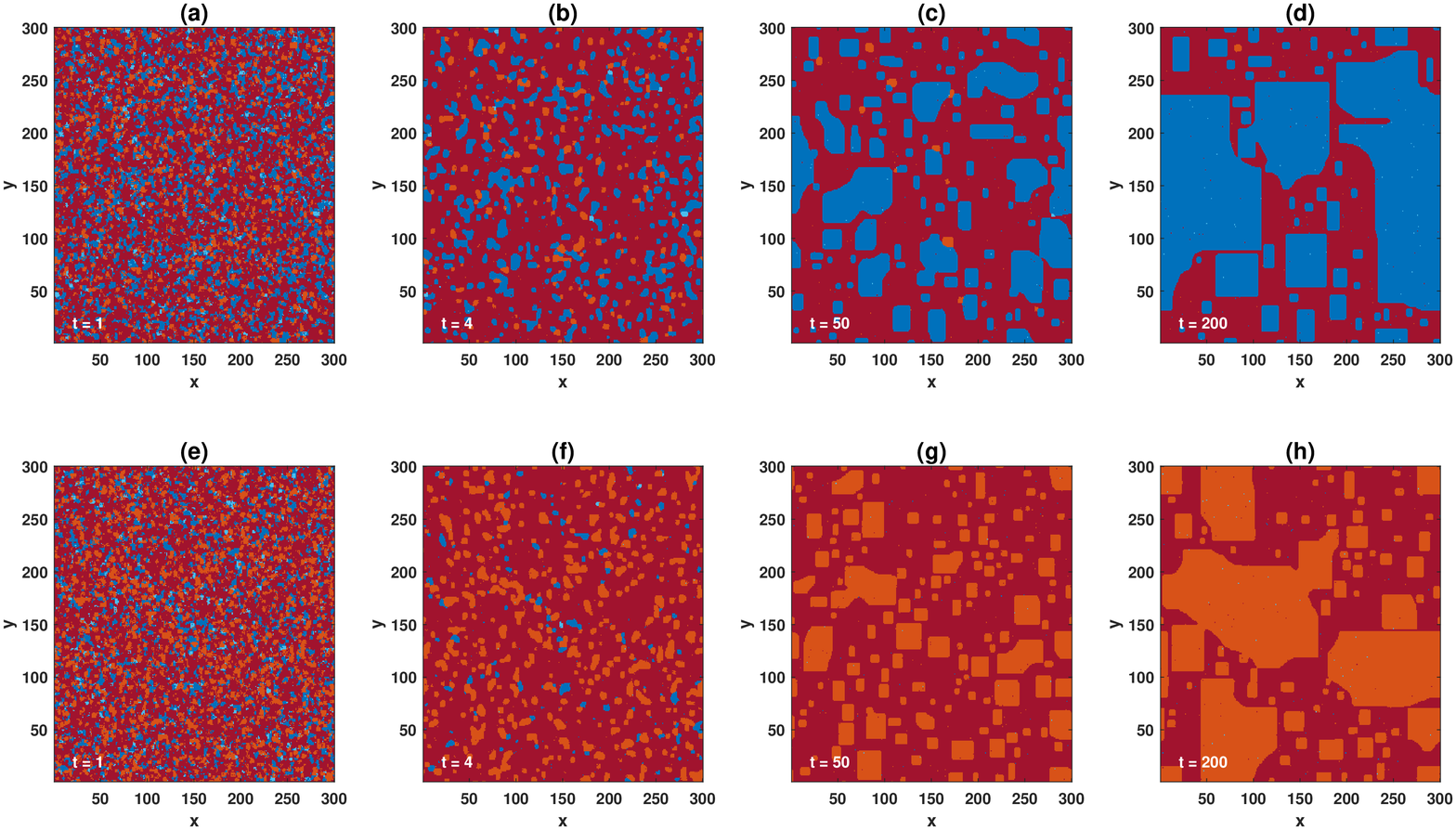}
	\caption{Time evolution of the system, close to the $PD-PC$ phase transition ((a) to (d)), and close to the $D-PD$ phase transition ((e) to (h)). Different strategies are indicated by different colors. Light blue shows non-punishing cooperators, dark blue shows punishing cooperators, light red shows non-punishing defectors, and dark red shows punishing defectors. (a) to (d): Starting from a random initial condition, close to the $PD-PC$ transition, punishing defectors rapidly increase in density and dominate the population. While all the other strategies go to extinction, punishing cooperators can survive by forming compact rectangular domains. The second stage of the evolution is governed by slow growth of punishing cooperators domains along their boundaries. (e) to (h): Close to the $D-PD$ transition, starting from a random initial condition, punishing defectors rapidly expand by driving all the other strategies into extinction. Only non-punishing defectors can survive by forming small rectangular domains. These blocks slowly increase in size in the see of punishing defectors along their boundaries, until they expand the whole population. In both cases, a population of $N=90000$ individuals lives on a two dimensional $300\times300$ lattice with periodic boundaries and Moore connectivity. In (a) to (d) $r=1.9$ and $\rho=5$, and in (e) to (h), $r=1.2$ and $\rho=4$. In all the cases $\nu=10^{-3}$, and $c=c'=1$.}
	\label{figvideoDPDandPDPC}
\end{figure}
\subsection{Growth process}
We end this section, by taking a deeper look into the growth process of different strategies close to the discontinuous transitions. To this goal, in Fig. (\ref{figtime}), we present the time evolution of the density of different strategies, close to the $D-PC$ transition. Here, $\nu=0.005$, and the population resides on a first nearest neighbor $100\times100$ two dimensional lattice with Moore connectivity and periodic boundaries. In (Fig. (\ref{figtime}.a), $r=1.72$ and $\rho=5$, These parameters are chosen such that the system is close to the $PC-PD$ transition, but in the $PC$ phase. In Fig. (\ref{figtime}.b), $r=1.2$ and $\rho=4.65$. These parameters are chosen such that the system is close to the $D-PD$ transition, but in the $D$ phase. As can be seen, in both cases the time evolution of the system shows two different time scales: a slow growth intermittent with short periods of fast growth. This gives the resulting time series a staircase-like appearance. To see why this is the case, in Fig. (\ref{figvideoCDandCCP}), Fig. (\ref{figvideoDPC}), and Fig. (\ref{figvideoDPDandPDPC}), we present snapshots of the time evolution of the system close to different phase transition. In Fig. (\ref{figvideoCDandCCP}.e) to Fig. (\ref{figvideoCDandCCP}.h), snapshots of the time evolution close to the $C-PC$ phase transition are represented, in Fig. (\ref{figvideoDPC}.a) to Fig. (\ref{figvideoDPC}.d), and also in Fig. (\ref{figvideoDPC}.e) to Fig. (\ref{figvideoDPC}.h), snapshots of the time evolution of two simulations for different parameter values, both chosen close to the $D-PC$ phase transition are presented, and in Fig. (\ref{figvideoDPDandPDPC}.a) to Fig. (\ref{figvideoDPDandPDPC}.d), snapshots of the time evolution of the system close to the $D-PD$ phase transition are presented, and finally, in Fig. (\ref{figvideoDPDandPDPC}.e) to Fig. (\ref{figvideoDPDandPDPC}.h), snapshots of the time evolution of the system close to the $PD-PC$ phase transition are presented. In all the cases, the population lives on a first nearest neighbor two dimensional $300\times300$ lattice with periodic boundaries and Moore connectivity. See the Figures for more detail.

As can be seen in the figures, in all the cases, close to the discontinuous transition, the dynamics involves (at least) two stages. Starting from a random initial condition, at the first stage of the evolution, one of the strategies rapidly grows and forms a sea of homogeneous strategies. Usually, one (in some cases more) of the strategies survive by forming compact rectangular blocks. The second stage of the evolution begins by the slow growth of these rectangular domains along their horizontal and vertical boundaries. The long periods of semi-stasis observed in Fig. (\ref{figtime}), results from such long periods of slow growth. However, at some time instances, two or more rectangular blocks meet, resulting in a non-rectangular island of individuals with the same growing strategy, in a sea of individuals with the same, but different strategy. Interestingly, the growth of such non-rectangular shapes posses a much faster time scale, resulting in the short periods of rapid growth observed in Fig. (\ref{figtime}).

Finally, for some parameter values, a rock-paper-scissor like dynamics is observed. An example is presented in Fig. (\ref{figvideoDPC}.a) to Fig. (\ref{figvideoDPC}.d), and in the Supplementary Video 3. In this case, starting from a random initial condition, punishing defectors rapidly grow and form a sea of punishing defectors, in which other strategies can only survive by forming small islands of homogeneous strategies. While punishing defectors can drive blocks of non-punishing and punishing cooperators into extinction, they are dominated by growing blocks of non-punishing defectors. These later blocks start invading the sea of punishing defectors until driving them to extinction. Although punishing defectors could have beaten punishing cooperators, this is not the case for non-punishing defectors. Once non-punishing defectors wash out punishing defectors, blocks of punishing cooperators surrounded by non-punishing defectors can grow until driving non-punishing defectors into extinction.

Another example of such rock-paper-scissor like dynamics occurs close to the $C-PC$ transition, Fig. (\ref{figvideoCDandCCP}.e) to Fig. (\ref{figvideoCDandCCP}.h) and Supplementary video 1. Here, punishing cooperators pave the way for the invasion of non-punishing cooperators, by driving defectors into extinction. When immune from defectors, blocks of non-punishing cooperators can outperform and grow in the sea of punishing cooperators. Non-punishing defectors, while in disadvantage in the presence of punishing cooperators, reappear in the system, once punishing cooperators get eliminated by non-punishing cooperators. We note that, this rock-paper-scissor like dynamic can facilitate the evolution of cooperation, due to elimination of defectors by punishing cooperators, close to the $C-D$ transition. Consequently, in our model, the $C-D$ transition shifts to smaller values of enhancement factors, $r$, compared to a system with the same network structure and size in a model where punishing strategies are absent. 

\section{Supplementary Videos}
In the Supplementary Videos (SV), the time evolution of the system, for a population of size $N=90000$ residing on a 2-dimensional $300\times300$ first nearest neighbor square lattice with Moore connectivity and periodic boundaries is presented. In the videos, we fix $\nu=10^{-3}$ and $\alpha=0.5$, and change the values of $r$ and $\rho$, such that the system is posed close to different phase transitions.

In SV.1 we have chosen $r=4$ and $\rho=3.47$. With these values, the system is in the $C$ phase, close to the $C-PC$ phase transition. As can be seen in the video, starting from a random initial condition, the defective strategies, $D$ and $PD$, are eliminated rapidly by punishing cooperators ($PC$). However, small rectangular-shape islands of non-punishing cooperators ($C$) are formed in the sea of punishing cooperators. As here, the parameters of the model are chosen such that the system is in the $C$ phase, non-punishing cooperators are in advantage with respect to punishing cooperators. Consequently, the small islands of non-punishing cooperators start to grow along the horizontal and vertical boundaries, until they invade the whole population. Non-punishing defectors, although unable to survive in the sea of punishing cooperators, can survive in the islands of non-punishing cooperators by forming narrow bands. Consequently, the density of non-punishing defectors increases as well, with increasing the density of non-punishing cooperators.

In the SV.2, we have set $r=2.8$, and $\rho=1.4$. With these values, the system is close to the $C-D$ transition, in the $C$ phase. As can be seen in the Video, starting from a random initial condition, non-punishing defectors rapidly drive other strategies into extinction, as here, the return to investments in punishing pool is not strong enough to promote punishing strategies. However, although they rapidly go to extinction, but by helping small islands of non-punishing cooperators to be formed in a sea of non-punishing defectors, they leave their footprint on the time evolution of the system. Afterwards, the dynamics of the system is driven by growth and collapse of such non-punishing cooperator islands in the sea of non-punishing defectors. We note that, in the absence of punishing strategies, non-punishing cooperators would have been unable to form small islands starting from a random initial condition, and would have been unable to survive, for this value of $r$ and network size. In this way, the very existence of punishing strategies can help the evolution of cooperation, even in a phase where such strategies can not survive.

In the SV.3 we present the dynamics of the system close to the $D-PC$ transition. Here, $r=2.2$ and $\rho=3.2$. As can be seen in the Video, starting from a random initial condition, domains of homogeneous strategies are formed rapidly. Interestingly, domains of non-punishing cooperators, can only survive if engulfed with punishing defectors, but rapidly go to extinction if in contact with non-punishing defectors. This is due to the fact that, as punishment enhancement factor is not large enough here, punishment imposes a large cost on the punishers which is not compensated by its return. This increases the fitness of non-punishing defectors with respect to punishing defectors. At the second stage of the evolution, the sea of non-punishing defectors grows and drives both the domains of punishing defectors and non-punishing cooperators into extinction. Only punishing cooperators, are in a high enough advantage with respect to non-punishing defectors to be able to survive in rectangular domains. This sets the stage for the third stage of the evolution, in which punishing cooperators begin to slowly grow along the horizontal and vertical boundaries. Interestingly, when two or more blocks come into contact due to slow growth along their boundaries, the domain of punishing cooperators rapidly grows until it takes a rectangular shape once again. This phenomenon results from the fact that, the growth of punishing cooperators is slowed along a horizontal or vertical boundary due to the fact that the amount of punishment a non-punishing defector receives from its neighboring punishing cooperators nearly offsets the advantage it receives due to not contributing to the public pool. However, as the number of neighboring punishing cooperators of a non-punishing defector increases, as it happens when the boundary of cooperators' domains is not a horizontal or vertical line, the amount of punishment a defector receives from punishing cooperators overcomes the gain the defector receives due to not contributing to the public good, by a large margin. Consequently, the speed of growth of punishing cooperators increases along such boundaries.

As mentioned before, the growth pattern described in the preceding paragraph possesses two different time scales. A slow time scale due to slow growth along the horizontal or vertical boundaries, and a fast time scale, due to rapid growth along other boundaries. As shown before, these two time scales lead to a staircase appearance of the time series of densities of different strategies, in which short periods of rapid growth are intermittent between long time intervals of slow growth. Such a phenomenology, is generally observed close to all the first order transitions of the system.

In SV.4 we present the dynamics of the system close to the $D-PD$ transition. Here, $r=1.2$ ‌and $\rho=4$. These values are chosen such that the system is in the $D$ phase, close to the $D-PD$ phase transition. As can be seen in the video, punishing defectors rapidly drive all the strategies into extinction. However, small domains of non-punishing defectors are formed in a sea of punishing defectors. As was the case, for other transitions, these blocks of non-punishing defectors start to grow along the boundaries. Furthermore, their growth shows tow different time scales. A small growth along the horizontal and vertical boundaries, intermittent with a rapid growth when different blocks collide.

Finally, in SV.5, we have set $r=1.9$ and $\rho=5$, such that the system is posed close to the $PC-PD$ transition, in the $PC$ phase. the same growth pattern observed in the other cases is at work here. Punishing defectors rapidly drive non-punishing defectors and non-punishing cooperators into extinction. Consequently, a sea of punishing defectors is formed in which all the other strategies can survive by forming homogeneous domains. Among these, only punishing cooperators are in a large enough advantage with respect to punishing defectors, to grow in the sea. Consequently, domains of punishing cooperators start to grow until they take over the whole population. As before, here the growth pattern of punishing cooperators show two different time scales: a slow growth along the horizontal and vertical boundaries, intermittent by short periods of fast growth when rectangular blocks of punishing cooperators collide, resulting in non-horizontal and non-vertical boundaries.

\begin{figure}%[!ht]
	\includegraphics[width=\linewidth, trim = 61 245 64 41, clip,]{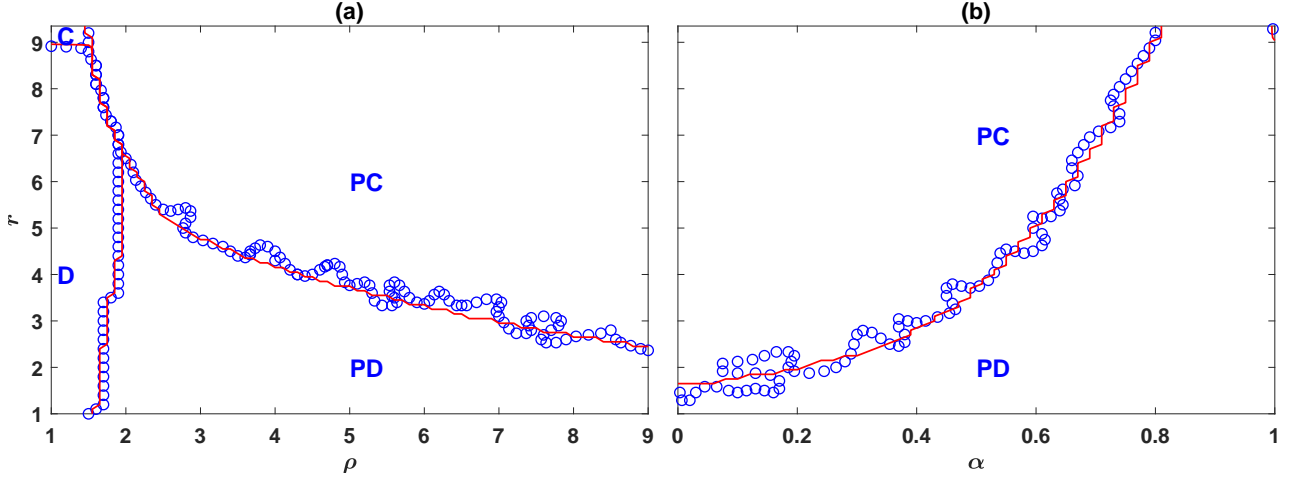}
	\caption{The phase diagram of the non-wasteful model for a mixed population. Blue circles denote the results of a simulation in a population of size $N=40000$, and the red lines denote the results of the replicator dynamics. Depending on the parameters of the model, the model shows four different phases separated with discontinuous transitions. $C$, $D$, $PC$, and $PD$ denote different phases in which, respectively, non-punishing cooperators, non-punishing defectors, punishing cooperators, and punishing defectors dominate the population. Here, $g=9$, $\nu=0.001$, and $c=c'=1$}
	\label{figadaptive}
\end{figure}

\section{The non-wasteful punishment model}
\subsection{non-wasteful punishment in a mixed population}
The phase diagram of the non-wasteful punishment model for a mixed population is presented in Fig. (\ref{figadaptive}). Fig. (\ref{figadaptive}.a), presents the phase diagram of the model in the $r-\rho$ plane, and Fig. (\ref{figadaptive}.b), presents the phase diagram of the model in the $r-\alpha$ plane. Blue circles represent the result of a simulation in a population of size $N=40000$, and the red lines represent the result of the replicator dynamics. As can be seen, the result of the replicator dynamics are in good agreement with the result of simulations.

As in the wasteful punishment model, the dynamics in the non-wasteful punishment model is multi-stable: depending on the initial conditions, the dynamics settle into a phase where one of the strategies dominates the population and drives all the other strategies into extinction. For small punishing enhancement factors, $\rho$, punishing strategies do not evolve. In this region, for $r$ smaller than a value close to $g$, the dynamics settle into a defective phase in which non-punishing defectors dominate the population. This phase is denoted by $D$ in the figure. On the other hand, for $r$ larger than $\sim g$, cooperators survive and dominate the population. This phase is denoted by $C$ in the figure. As $\rho$ increases, a phase transition to a phase where punishing strategies evolve occurs. However, the nature of the evolved punishing strategy depends on the value of $r$. For small $r$, such that the return to the investment in the public good is small, punishing defectors dominate the population. This phase is denoted by $PD$ in the figure. On the other hand, for large values of $r$, the dynamics settle into a phase where punishing cooperators dominate the population. This phase is denoted by $PC$ in the figure. 

\begin{figure}%[!ht]
	\includegraphics[width=\linewidth, trim = 55 189 60 30, clip,]{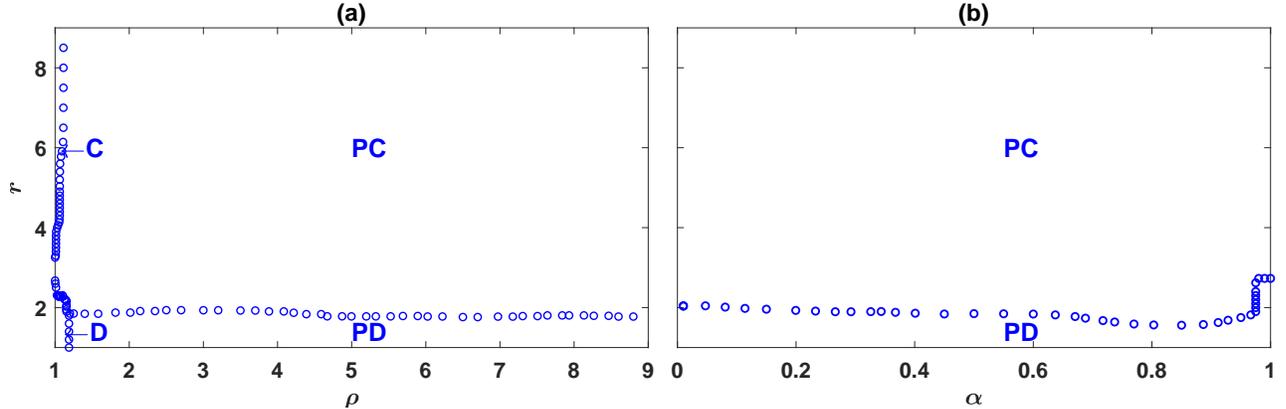}
	\caption{The phase diagram of the non-wasteful model for a structured population. The phase diagram is derived by running simulations in a population of size $160000$, residing on a $400\times 400$ first nearest neighbor lattice with Moore connectivity and periodic boundaries. Depending on the parameters of the model, the model shows four different phases. $C$, $D$, $PC$, and $PD$ denote different phases in which, respectively, only, cooperators, defectors, punishing cooperators, and punishing defectors survive.  Here, $g=9$, $\nu=0.001$, and $c=c'=1$. In (a) $\alpha=0.5$ and in (b) $\rho=5$.}
	\label{fignetadaptive}
\end{figure}

Comparison with the wasteful punishment model shows that the evolution of punishment is facilitated in the non-wasteful punishment model. This can be seen by noting that the phase transition to the punishing phase occurs for a smaller value of punishment enhancement factor, $\rho$, in the non-wasteful punishment model. This shows a smaller return to the investment in the punishment pool is sufficient to give rise to the evolution of punishment, when, instead of being wasted, the resources in the punishment pool are redistributed among its contributors in case there is nobody to punish, .

We note that, similarly to the wasteful punishment model, higher returns to the investments in the punishment pool facilitate the evolution of social as opposed to the anti-socail punishment. This can be seen by noting that for higher values of $\rho$, the transition to the social punishment phase shifts to smaller values of $r$. That is, for more effective punishment mechanisms, a smaller enhancement factor for the public resource is sufficient to promote social punishment.

The phase diagram of the non-wasteful model in the $\alpha-r$ plane is presented in Fig. (\ref{figadaptive}.b). As can be seen, in the non-wasteful model punishing strategies evolve even for $\alpha=0$. That is, punishment evolves even in the absence of second-order punishment. The reason is that, the prospect of receiving return from the punishment pool in case there is nobody to punish, can act as a reward which solves the second-order free-riding problem. Consequently, second-order punishment is not necessary to ensure the evolution of punishment. Furthermore, the value of $r$ for which social punishment evolves increases by increasing $\alpha$. This shows second-order punishment, counter intuitively, is detrimental for the evolution of social punishment in a situation where the resources of punishment institute are not wasted when there are nobody to punish.

Finally, we note that, similarly to the wasteful punishment model, the non-wasteful punishment model shows multistability in the whole region of the phase diagram: For $r<g$, there exist three stable phases, $D$ where non-punishing defectors dominate, $PD$, where punishing defectors dominate, and $PC$, where punishing cooperators dominate. Depending on the initial conditions, the dynamics settle into one of these phases. On the other hand, for $r>g$, both the $C$ phase, in which non-punishing cooperators dominate the population, and the $PC$ phase, where punishing cooperators dominate the population are stable. This implies that the phase transitions involving different punishment phases in this model are discontinuous.

\subsection{Non-wasteful punishment in a structured population}
We present the phase diagram of the non-wasteful punishment model in the case of a structured population in Fig. (\ref{fignetadaptive}). Here, simulations are performed in a population of size $N=160000$ individuals residing on a $400\times 400$ first nearest neighbor two dimensional square lattice, with Moore connectivity and periodic boundaries. In Fig. (\ref{fignetadaptive}.a) we have set $g=9$, $\nu=10^{-3}$, and $\alpha=0.5$, and in Fig. (\ref{fignetadaptive}.b), we have set $g=9$, $\nu=10^{-3}$, and $\rho=5$.

The non-wasteful punishment model in a structured population shows similar phases to those observed in the wasteful punishment model. However, there are interesting differences with the wasteful punishment model. First, for a fixed $r$, the phase transition to a phase where punishing institutions evolve shifts to smaller punishment enhancement factors. In contrast, the transition from antisocial punishment to social punishment for a fixed $\rho$ does not show significant difference in the non-wasteful punishment model, compared to the wasteful punishment model. We note that, the transition to the evolution of punishing institution occurs for a smaller value of $\rho$ in a structured population, compared to that in the case of a well-mixed population. This shows that when punishment is non-wasteful, it evolves with more ease in a structured population compared to a mixed population. This contrasts the situation in the wasteful punishment model where, as we saw, network structure could hinder the evolution of punishing institutions. 

Interestingly, the transition from the $D$ phase to the $C$ phase is lost in the non-wasteful punishment model. Instead, by increasing $r$ for a small $\rho$, the system shows a phase transition from the $D$ phase to the $PC$ phase. Further increasing $r$, the system shows a phase transition from the $PC$ phase to the $C$ phase. This phenomenology suggest that, just like the wasteful punishment model, the evolution of social punishment is facilitated close to the $C-D$ phase transition as well, as in this region, social punishment evolves for smaller returns to the investments in the punishment pool, $\rho$.

The phase diagram in the $\alpha-r$ plane shows that, contrary to the wasteful punishment model, second order punishment is not necessary for the evolution of punishment in this case: For the chosen value of $\rho$, punishing institutes evolve for all the values of $\alpha$. For smaller $r$, antisocial punishment evolves, as $r$ increases, social punishment evolves. As we saw, this was the case in a well-mixed population as well.

\end{document}